\documentclass[10pt,Journal]{IEEEtran}\IEEEoverridecommandlockouts
\usepackage{algorithmic}
\usepackage[linesnumbered,ruled]{algorithm2e}
\usepackage{amsfonts}
\usepackage{amsmath}
\usepackage{amssymb}
\usepackage{amsthm}
\usepackage{bm}
\usepackage{bbm}
\usepackage{booktabs}
\usepackage{caption}
\usepackage{CJK}
\usepackage{color}
\usepackage{cuted}
\usepackage{geometry}
\usepackage{graphicx}
\usepackage{indentfirst}
\usepackage[marginal]{footmisc}
\usepackage{multirow}
\usepackage{makecell}
\usepackage{rotating}
\usepackage{subcaption}

\allowdisplaybreaks[4]

\newenvironment{shrinkeq}[1]%
{\bgroup
 \addtolength\abovedisplayshortskip{#1}
 \addtolength\abovedisplayskip{#1}
 \addtolength\belowdisplayshortskip{#1}
 \addtolength\belowdisplayskip{#1}}
{\egroup\ignorespacesafterend}

\theoremstyle{plain}
\newtheorem{thm}{Theorem}

\newtheorem{prop}{Proposition}
\newtheorem{cor}{Corollary}

\begin{document}

\title{\huge{Rethinking Soft Interference Cancellation (IC) for MIMO: A Hard-Decision IC Inspired Recursive Scheme}}

\author{Yanze Zhu, Hufei Zhu, Qingqing Wu, Yikui Zhai, Wen Chen, and Yang Liu \thanks{\hspace{0.35cm}Y. Zhu, Q. Wu, and W. Chen are with the Department of Electronic Engineering, Shanghai Jiao Tong University, 200240, China, email: \{yanzezhu, qingqingwu, wenchen\}@sjtu.edu.cn.}\thanks{\hspace{0.35cm}H. Zhu and Y. Zhai are with the School of Electronics and Information Engineering, Wuyi University, Jiangmen 529020, Guangdong, China, email: hufeizhu93@hotmail.com, yikuizhai@163.com.}\thanks{\hspace{0.35cm}Y. Liu is with the School of Information and Communication Engineering, Dalian University of Technology, Dalian, China, and also with the National Mobile Communications Research Laboratory, Southeast University, Nanjing, China, email: yangliu\_613@dlut.edu.cn.}\thanks{\hspace{0.35cm}Y. Zhu and H. Zhu contributed equally to this work.}
\vspace{-7mm}}%\thanks{\hspace{0.35cm}Corresponding authors: Qingqing Wu and Yang Liu.}

\maketitle

\begin{abstract}
Multiple-input multiple-output (MIMO) technology has been regarded as one of the most important technologies to enable emerging applications in current and next generation wireless communication systems, for which signal detection methods have been endowed with higher requirements, such as finer bit-error ratio (BER) performance, lower complexity, and smaller memory. Existing detectors mainly include hard-decision-based ordered successive interference cancellation (HD-OSIC) schemes with relatively simple implementation, and linear-minimum-mean-square-error-based iterative soft interference cancellation (LMMSE-ISIC) schemes exhibiting near-optimal BER performance, whose advantages are combined by the detector developed in this paper. Specifically, we first elaborate that the LMMSE-ISIC scheme is the extension of the HD-OSIC counterpart, via comparing our proposed reordered description based on the equivalent channel matrix for the LMMSE-ISIC detection process with the other. Then, we propose a recursive scheme with speed advantage and memory saving for LMMSE-ISIC by extending that for HD-OSIC, where the LMMSE-ISIC estimate and the filtering bias are updated highly efficiently. Compared to the existing best low-complexity LMMSE-ISIC scheme, theoretically, the required computations and memory units in each iteration of our proposed scheme decrease by at least 87.50\% and 80.00\%, respectively, and simulation results demonstrate that our proposed scheme always yields identical BER performance.
\end{abstract}

\vspace{-1.5mm}

\vspace{-3mm}
\section{Introduction}

Multiple-input multiple-output (MIMO) technology, which can exploit the extra spatial degree-of-freedom (DoF) in rich multi-path environments to achieve significantly high spectral efficiency [1], has been envisioned as one of the critical technologies in the sixth-generation (6G) wireless communication systems by both academia and industry.

Among the challenges of MIMO, signal detection is a fundamental problem [2] that affects the implementation of 6G remarkably. Compared to the fifth-generation (5G) systems, 6G systems require extremely-high data rate, ultra-low latency, dozens of times higher energy efficiency (EE), lower hardware cost, and so on [3], which put forward new requirements for detection methods. For one thing, the bit-error ratio (BER) performance yielded by detection schemes should be better to satisfy the 6G requirements of 1 Tb/s peak data rate and 99.99999\% reliability [4]. For another, the detection methods should have lower computational complexity to achieve the latency of 0.1 millisecond (ms) [4]. On the other hand, by reducing complexity, the power consumption can be reduced, while decreasing the quantization bit width of analog-to-digital converter (ADC) and the number of radio frequency (RF)-chains to save hardware cost. In addition, the memory size utilized by detection schemes should be smaller. The reason lies in that, according to [5], the power consumption for “moving data” in 1-bit floating-point can be tens of times higher than that for “computing data”. Via leveraging detection methods with smaller memory requirements, the number of off-chip memory accesses can be significantly reduced for power saving, supporting the goal of EE improvement in 6G systems. Furthermore, a smaller-memory scheme can allow it to reside in on-chip static random access memory (SRAM) such that additional external dynamic RAM (DRAM) or flash memory is indeed unnecessary for chip area shrinkage and cost compression. Besides, the available memory of most Internet-of-Things (IoT) devices is limited [6], which indicates that the detection methods requiring smaller memories are suitable for implementing most terminals. Hence, exploring novel signal detection schemes with finer BER performance while saving computations and memories is essential for enabling 6G scenarios, e.g., immersive communication, massive communication, hyper reliable and low-latency communication, and so on [7].

Considering a MIMO system where $N$ signals are simultaneously transmitted from the transmitter equipped with $N$ antennas to the receiver equipped with $M$ ($\geq N$) antennas, the signal detection aims at accurately recovering these $N$ transmit signals. In the existing literature, various signal detection methods have been proposed. For instance, in [8]-[10], the maximum-likelihood (ML) detectors were proposed, which can yield optimal BER performance. However, such detectors will incur excessive computational complexity that increases exponentially with $N$, which are difficult to implement in practice, especially in 6G systems where $N$ can reach up to hundreds or even thousands [4]. To overcome this drawback, the hard-decision-based ordered successive interference cancellation (HD-OSIC) detector [9] was proposed, which detects $N$ transmit signals sequentially in an optimal order based on signal-to-noise ratio (SNR). Specifically, the transmit signal with the highest SNR is first detected through a linear zero-forcing (ZF) or minimum mean square error (MMSE) filtering vector, and then the hard decision of the detected symbol is utilized to eliminate the corresponding interference in the receive signal vector. The above procedure repeats with the next strongest signal among all the undetected signals until all $N$ signals have been detected. Compared to the ML detector, the HD-OSIC detector significantly reduces the computational complexity with slight BER performance degradation. Hence, the HD-OSIC scheme has been widely adopted in practical use, e.g., the vertical Bell Laboratories Layered Space-Time (V-BLAST) architecture [11]. When computing the filtering vectors, the conventional HD-OSIC detector needs to compute $N$ matrix inversions for the pseudoinverses of the deflated channel matrices [11], which leads to high computational complexity. To tackle this issue, efficient schemes including recursive schemes [12]-[16] and the square-root schemes [17]-[20] have been proposed, which update the detection error covariance matrix and its square-root matrix, respectively, to avoid most matrix inversions in the conventional detector.

Another popular kind of detector for MIMO is the so-called linear-MMSE-based iterative soft interference cancellation (LMMSE-ISIC) detector [21]-[24]. The LMMSE-ISIC detector is an iterative near-optimal scheme consisting of $K$ ($\geq 1$) iterations, in each of which all $N$ transmit signals are detected sequentially in an order decided arbitrarily or based on SNR. When detecting a signal, the soft decisions of all the other $(N - 1)$ signals are utilized to subtract their corresponding interferences in the receive signal vector, followed by computing the LMMSE-ISIC estimate of that signal through the filtering vector considering the residual interference (after soft interference cancellation) from the remaining $(N - 1)$ signals. In the conventional LMMSE-ISIC scheme, the computation of the filtering vectors depends on $N$ inversions of the $M \times M$ covariance matrix during each of the $K$ iterations [23], [24]. To reduce the complexity of the LMMSE-ISIC detector caused by frequent inverse matrix computations, several solutions have been proposed in the existing literature [24]-[28]. For instance, the authors of [24] designed a low-complexity LMMSE-ISIC scheme, which updates the $N \times N$ inverse matrix rather than most matrix inversions of relatively larger size $M \times M$ in the conventional LMMSE-ISIC detector. The work [25] developed a low-complexity LMMSE-ISIC scheme by leveraging Neumann-based algorithm, which achieves significant complexity reduction. In [26], the Gauss-Seidel-based method was considered to realize the low-complexity LMMSE-ISIC scheme, which exhibits superior BER performance over the scheme proposed in [25] while decreasing the number of complex multiplications. Besides, the works [27] and [28] adopted the conjugate-gradient-based and Lanczos-based schemes in LMMSE-ISIC detector, respectively, to avoid computation-demanding high-dimensional matrix inversions.

Although the potentials of the HD-OSIC and LMMSE-ISIC detectors have been deeply explored by the existing literature, e.g., [9], [11]-[28], they both have inherent shortcomings, respectively. Specifically, the HD-OSIC scheme usually neglects the residual interference by simply assuming the perfect interference cancellation without errors, hence, the implementation of HD-OSIC detectors is simple. However, due to the fact that the detection error and the corresponding residual interference always exist [29], the BER performance yielded by HD-OSIC scheme is indeed insufficiently fine, especially when bit error and the associated error propagation appear. In contrast, the LMMSE-ISIC counterpart takes the residual interference into consideration, therefore, the LMMSE-ISIC detectors can exhibit near-optimal BER performance, at the cost of more complicated implementation.

Based on the above discussions, we naturally raise a question: \emph{Can we combine the individual advantages of the HD-OSIC and LMMSE-ISIC detectors} to achieve a near-optimal detector with adequately low complexity? Before answering this question, it is meaningful to clarify \emph{whether there exists a certain relationship between both detectors}. Once it exists, we can take advantage of this relationship by directly extending the procedure of one scheme to the other. Unfortunately, to the best of the authors’ knowledge, the above two questions are still open problems, which have never been considered in the existing literature, e.g., [9], [11]-[28]. These unresolved problems motivate this work, whose solutions are successfully developed. The contributions of this paper are stated as follows.

\begin{table}[!t]
\centering
\caption{Comparison of Complexity and Memory for Existing Low-Complexity LMMSE-ISIC Schemes.}
\begin{tabular}{|c|c|c|c|}\hline
Reference & Proposed Scheme & Complexity & Memory\\\hline
[25] & Neumann & High & Large\\\hline
[26] & Gauss-Seidel & Modest & Modest\\\hline
[27] & Conjugate-Gradient & Modest & Mosest\\\hline
[28] & Lanczos & Modest & Modest\\\hline
\makecell[c]{[24] \\ (Existing Best)} & AMMSE & Low & Small\\\hline
\textbf{This work} & \textbf{Recursive} & \makecell[c]{\textbf{Very Low} \\ ($\mathbf{\leq}$\textbf{1/8 of [24]})} & \makecell[c]{\textbf{Very Small} \\ ($\mathbf{\leq}$\textbf{1/5 of [24]})}\\\hline
\end{tabular}
\end{table}

\begin{itemize}

\item First, we propose to utilize the concept of the equivalent channel matrix, which is related to the channel matrix and the residual interference variance matrix, to innovatively present a reordered description of the detection process for LMMSE-ISIC. Then, by comparing this novel reordered description for LMMSE-ISIC with HD-OSIC detection process, we discover that the former is the extension of the latter. To the best of the authors’ knowledge, this amazing conclusion is for the first time obtained by this paper, which has never appeared in the existing literature, e.g., [9], [11]-[28].

\item Second, we propose to extend the efficient recursive scheme for HD-OSIC detection process to LMMSE-ISIC counterpart, aiming at developing a detector with sufficiently fine BER performance, low complexity, and small memory requirement. Specifically, the computations of the LMMSE-ISIC estimate and the filtering bias are achieved by utilizing the Hermitian inverse matrix and the uniform symbol estimate vector, which are updated highly efficiently. Note that existing works have never considered to design such a detector combining the advantages of HD-OSIC and those of LMMSE-ISIC, e.g., [9], [11]-[28]. To the best of the authors’ knowledge, this is the first work extending an efficient detector for HD-OSIC to LMMSE-ISIC.

\item Third, we theoretically analyze the complexity and memory requirement of our proposed recursive LMMSE-ISIC scheme, which is compared with the scheme proposed in [24] with the lowest complexity and smallest memory requirement among the existing low-complexity LMMSE-ISIC schemes. It can be observed that, the required computations and memory units of our proposed scheme is less than $\frac{1}{8}$ and $\frac{1}{5}$ of those required by the scheme in [24], respectively. As shown in Table I, we make a brief comparison between our proposed solution and existing low-complexity LMMSE-ISIC schemes to emphasize that all the existing low-complexity LMMSE-ISIC schemes have much higher complexities and needs much larger memories than ours, e.g., [24]-[28].

\item Last but not least, we provide extensive simulation results to verify the unique superiority of our proposed recursive low-complexity LMMSE-ISIC detector. For one thing, the effectiveness of theoretical analysis is demonstrated via simulations, where the practical speedup of computational complexity compared to the scheme developed in [24] approaches the maximum with $M$ modestly large. For another, the BER performance yielded by our proposed recursive low-complexity LMMSE-ISIC scheme coincides that achieved by the scheme in [24], however, the computational complexity and memory requirement of the former are significantly much lower than the latter.

\end{itemize}

The rest of this paper is organized as follows. Section II introduces the system model for MIMO and the preliminaries of HD-OSIC and LMMSE-ISIC. Section III elaborates that the LMMSE-ISIC detection process is the extension of the HD-OSIC counterpart. Section IV develops a recursive low-complexity scheme for LMMSE-ISIC and analyzes its computational complexity and memory requirement. Simulation results are provided in Section V to verify the advantages of our proposed recursive low-complexity LMMSE-ISIC scheme. The conclusion of this paper is drawn in Section VI.

\emph{Notations:} $\mathbf{X}^{\mathsf{T}}$, $\mathbf{X}^{\mathsf{H}}$, and $\mathbf{X}^{-1}$ denote the transposition, conjugate transposition, and inversion of matrices, respectively. $\mathbf{I}_{n}$ and $\bm{0}_{n}$ denote the $n \times n$ identity matrix and the $n \times 1$ zero vector, respectively. $[\mathbf{X}]_{:,n}$ and $[\mathbf{X}]_{n,:}$ denote the $n$-th column and row of the matrix $\mathbf{X}$, respectively. $[\mathbf{x}]_{n}$ and $[\mathbf{X}]_{n,n}$ denote the $n$-th entry and the $n$-th diagonal entry of the vector $\mathbf{x}$ and the square matrix $\mathbf{X}$, respectively. $\mathbb{E} \{ \cdot \}$ denotes the mathematical expectation. $\mathbf{A} = \mathsf{Diag}(\mathbf{a})$ denotes the diagonal matrix $\mathbf{A}$ with the entries of the vector $\mathbf{a}$ on the main diagonal.

\vspace{-0.5mm}
\section{System Model and Existing Approaches}

In this section, we first describe the system model for MIMO. Then, we introduce the conventional HD-OSIC detection process [11] and the corresponding recursive implementation [16] with the least computations and memories. Besides, we present the conventional LMMSE-ISIC detection process [21]-[23] and the corresponding recently proposed low-complexity implementation [24].

\vspace{-1mm}
\subsection{System Model}

We consider a spatially multiplexed MIMO system where the transmitter and receiver are equipped with $N$ and $M(\geq N)$ antennas, respectively. Denote the transmit signal vector as $\mathbf{x} = [x_{1}, \cdots, x_{N}]^{\mathsf{T}} \in \mathbb{C}^{N \times 1}$, where $x_{n}$ ($n \in \mathcal{N} \triangleq \{ 1, \cdots, N \}$) is the symbol transmitted from the $n$-th transmit antenna. Then, the receive signal vector $\mathbf{y} \in \mathbb{C}^{M \times 1}$ is given by
\begin{shrinkeq}{-0.5ex}
\begin{align}
\mathbf{y} = \mathbf{H}\mathbf{x} + \mathbf{n},
\end{align}
\end{shrinkeq}
where $\mathbf{H} = [\mathbf{h}_{1}, \cdots, \mathbf{h}_{N}] \in \mathbb{C}^{M \times N}$ represents the channel matrix with $\mathbf{h}_{n}$ ($n \in \mathcal{N}$) being the $n$-th column of $\mathbf{H}$$^{\text{1}}$\footnote{$^{\text{1}}$Note $\mathbf{H}$ can be obtained via existing channel estimation techniques [30], [31].}, and $\mathbf{n} \in \mathbb{C}^{M \times 1}$ represents the additive white Gaussian noise (AWGN) following the distribution $\mathcal{CN}(\bm{0}_{M}, \sigma^{2}\mathbf{I}_{M})$.

\vspace{-0mm}
\subsection{Preliminary of HD-OSIC}

\subsubsection{Conventional Scheme}

For the channel model (1), the conventional HD-OSIC detection process [11] includes $N$ procedures to detect $N$ entries of $\mathbf{x}$ iteratively with the optimal order based on SNR. In the $i$-th ($i \in \mathcal{N}$) procedure, the entry with the highest SNR among all the undetected $n = N - i + 1$ entries is permuted to be the $n$-th entry $x_{n}$ and estimated by
\begin{shrinkeq}{-0.5ex}
\begin{align}
\hat{x}_{n} = [(\mathbf{H}_{n}^{\mathsf{H}}\mathbf{H}_{n} + \sigma^{2}\mathbf{I}_{n})^{-1}\mathbf{H}_{n}^{\mathsf{H}}]_{n,:}\mathbf{y}^{(n)},
\end{align}
\end{shrinkeq}
where $\mathbf{H}_{n} = [\mathbf{h}_{1}, \cdots, \mathbf{h}_{n}]$ includes the first $n$ columns of the permuted $\mathbf{H}$, and $\mathbf{y}^{(n)}$ ($n = N, \cdots, 1$) represents the hard-interference (HI)-cancelled receive vector. Based on $\hat{x}_{n}$, we obtain the hard-decision $\check{x}_{n}$. Then, assuming perfect interference cancellation without errors, $x_{n}$ is cancelled from $\mathbf{y}^{(n)}$, which can be expressed as
\begin{shrinkeq}{-0.5ex}
\begin{align}
\mathbf{y}^{(n - 1)} = \mathbf{y}^{(n)} - \check{x}_{n}\mathbf{h}_{n}.
\end{align}
\end{shrinkeq}
Hence, by iteratively repeating the procedure with (2) and (3) for $n = N, \cdots, 1$ with the initial $\mathbf{y}^{(N)} = \mathbf{y}$, all entries of $\mathbf{x}$ can be detected.

\subsubsection{Efficient Recursive Scheme}

Note the above conventional HD-OSIC detection process utilizing (2) and (3) leads to high computational complexities and exceeding memory units. To tackle this issue, several efficient recursive schemes have been proposed in the existing literature, e.g., [12]-[16], and we will introduce the scheme [16] with the least computations and memories among existing works to the best of the authors' knowledge.

The efficient recursive schemes for HD-OSIC detection process are based on the $n \times n$ inverse matrix $\mathbf{Q}_{n} = (\mathbf{H}_{n}^{\mathsf{H}}\mathbf{H}_{n} + \sigma^{2}\mathbf{I}_{n})^{-1} = \mathbf{R}_{n}^{-1}$ [12]-[16], where $\mathbf{R}_{n} \triangleq \mathbf{H}_{n}^{\mathsf{H}}\mathbf{H}_{n} + \sigma^{2}\mathbf{I}_{n}$ can be partitioned by
\begin{shrinkeq}{-0.5ex}
\begin{align}
\mathbf{R}_{n} = \begin{bmatrix}
\mathbf{R}_{n - 1} & \ddot{\mathbf{r}}_{n}\\
\ddot{\mathbf{r}}_{n}^{\mathsf{H}} & \gamma_{n}\\
\end{bmatrix}
\end{align}
\end{shrinkeq}
with $\mathbf{R}_{n - 1} \triangleq \mathbf{H}_{n - 1}^{\mathsf{H}}\mathbf{H}_{n - 1} + \sigma^{2}\mathbf{I}_{n - 1}$. According to the lemma for inverting a partitioned matrix [32, Eq. 8], $\mathbf{Q}_{n}$ can be calculated as
\begin{shrinkeq}{-0.5ex}
\begin{align}
\mathbf{Q}_{n} = \begin{bmatrix}
\ddot{\mathbf{Q}}_{n - 1} & \ddot{\mathbf{q}}_{n}\\
\ddot{\mathbf{q}}_{n}^{\mathsf{H}} & \omega_{n}\\
\end{bmatrix}
\end{align}
\end{shrinkeq}
with
\begin{shrinkeq}{-0.5ex}
\begin{align}
\omega_{n} &= (\gamma_{n} - \ddot{\mathbf{r}}_{n}^{\mathsf{H}}\mathbf{Q}_{n - 1}\ddot{\mathbf{r}}_{n})^{-1}, \\
\ddot{\mathbf{q}}_{n} &= -\omega_{n}\mathbf{Q}_{n - 1}\ddot{\mathbf{r}}_{n}, \\
\ddot{\mathbf{Q}}_{n - 1} &= \mathbf{Q}_{n - 1} + \omega_{n}^{-1}\ddot{\mathbf{q}}_{n}\ddot{\mathbf{q}}_{n}^{\mathsf{H}}.
\end{align}
\end{shrinkeq}
In the \emph{initialization} phase, the initial $\mathbf{Q}_{N}$ can be obtained by utilizing the above (5)-(8) to expand $\mathbf{Q}_{n - 1}$ to $\mathbf{Q}_{n}$ iteratively for $n = 2, \cdots, N$. In the \emph{recursion} phase, $\mathbf{Q}_{n}$ (permuted according to the detection order) is deflated to $\mathbf{Q}_{n - 1}$ by $\mathbf{Q}_{n - 1} = \ddot{\mathbf{Q}}_{n - 1} - \omega_{n}^{-1}\ddot{\mathbf{q}}_{n}\ddot{\mathbf{q}}_{n}^{\mathsf{H}}$ [15], which can be derived from (8).

Define the symbol estimator vector as $\breve{\mathbf{t}}_{n} \triangleq \mathbf{Q}_{n}\mathbf{H}_{n}^{\mathsf{H}}\mathbf{y}^{(n)}$. Then, the estimate of $x_{n}$ can be obtained from $\breve{\mathbf{t}}_{n}$ by
\begin{shrinkeq}{-0.5ex}
\begin{align}
\hat{x}_{n} = [\breve{\mathbf{t}}_{n}]_{n},
\end{align}
\end{shrinkeq}
and the interference of $x_{n}$ can be cancelled equivalently by
\begin{shrinkeq}{-0.5ex}
\begin{align}
\breve{\mathbf{t}}_{n - 1} = \breve{\mathbf{t}}_{n}^{[-n]} + \omega_{n}^{-1}(\check{x}_{n} - [\breve{\mathbf{t}}_{n}]_{n})\ddot{\mathbf{q}}_{n},
\end{align}
\end{shrinkeq}
where $\breve{\mathbf{t}}_{n}^{[-n]}$ denotes the permuted $\breve{\mathbf{t}}_{n}$ with the last entry removed. By utilizing (9) and (10) instead of (2) and (3), respectively, the recursive HD-OSIC scheme can not only save computations, but also save memories since only the entries in $\mathbf{Q}_{n}$, i.e., $\ddot{\mathbf{q}}_{n}$ and $\omega_{n}$, are utilized.

\vspace{-2mm}
\subsection{Preliminary of LMMSE-ISIC}

\subsubsection{Conventional Scheme}

In each iteration of the conventional LMMSE-ISIC detection process [21]-[23], $N$ transmit symbols are estimated sequentially according to a detection order. Assume the index of the $n$-th ($n \in \mathcal{N}$) detected symbol to be just $n$ for simplicity, i.e., $\mathsf{o}(n) = n$, and denote $\mathsf{o}(n - i)$ as the index of the $i$-th last symbol detected before detecting $x_{n}$, i.e., [24]
\begin{shrinkeq}{-0.5ex}
\begin{align}
\mathsf{o}(n - i) = \begin{cases}
n - i + N, & \text{if } n - i \leq 0, \\
n - i, & \text{if } 1 \leq n - i \leq N.
\end{cases}
\end{align}
\end{shrinkeq}

For each $x_{n}$ ($n \in \mathcal{N}$), the LMMSE-ISIC detector computes the LMMSE-ISIC estimate $\hat{x}_{n}$ and the MMSE filtering bias $\mu_{n}$, and then utilizes $\hat{x}_{n}$ and $\mu_{n}$ to compute the soft decision $\bar{x}_{n}$ and the residual interference variance $v_{n}$, which are indeed the conditional symbol mean and variance of $x_{n}$, respectively.

The computations of the LMMSE-ISIC estimate $\hat{x}_{n}$ and the filtering bias $\mu_{n}$ are based on the soft decision vector $\bar{\mathbf{x}} = [\bar{x}_{1}, \cdots, \bar{x}_{N}]^{\mathsf{T}}$ and the residual interference variance matrix $\mathbf{V} = \mathsf{Diag}([v_{1}, \cdots, v_{N}]^{\mathsf{T}})$. In the procedure for $x_{n}$, the soft interference (SI)-cancelled receive vector $\tilde{\mathbf{y}}^{\hat{n}}$ for detecting $x_{n}$ is given by
\begin{shrinkeq}{-0.5ex}
\begin{align}
\tilde{\mathbf{y}}^{\hat{n}} = \mathbf{y} - \sum_{i=1, i \neq n}^{N}\mathbf{h}_{i}\bar{x}_{i} = \mathbf{y} - \mathbf{H}\bar{\mathbf{x}}^{\hat{n}},
\end{align}
\end{shrinkeq}
where $\bar{\mathbf{x}}^{\hat{n}} = [\bar{x}_{1}, \cdots, \bar{x}_{n - 1}, 0, \bar{x}_{n + 1}, \cdots, \bar{x}_{N}]^{\mathsf{T}}$ represents the soft decision vector $\bar{\mathbf{x}}$ with the $n$-th entry $\bar{x}_{n}$ being replaced by $0$. On the other hand, the filtering vector $\mathbf{f}_{\mathrm{CV},n}$ for $x_{n}$ can be expressed as
\begin{shrinkeq}{-0.5ex}
\begin{align}
\mathbf{f}_{\mathrm{CV},n} = (\mathbf{H}\mathbf{V}^{\hat{n}}\mathbf{H}^{\mathsf{H}} + \sigma^{2}\mathbf{I}_{M})^{-1}\mathbf{h}_{n} = \mathbf{D}^{\hat{n}}\mathbf{h}_{n},
\end{align}
\end{shrinkeq}
where $\mathbf{V}^{\hat{n}} = \mathsf{Diag}([v_{1}, \cdots, v_{n - 1}, 1, v_{n + 1}, \cdots, v_{N}]^{\mathsf{T}})$ represents the residual interference variance matrix $\mathbf{V}$ with the $n$-th diagonal entry $v_{n}$ being replaced by $1$, and $\mathbf{D}^{\hat{n}} \triangleq (\mathbf{H}\mathbf{V}^{\hat{n}}\mathbf{H}^{\mathsf{H}} + \sigma^{2}\mathbf{I}_{M})^{-1}$ represents the covariance matrix for $\tilde{\mathbf{y}}^{\hat{n}}$, i.e., $\mathbf{D}^{\hat{n}} = \mathbb{E} \{ \tilde{\mathbf{y}}^{\hat{n}}(\tilde{\mathbf{y}}^{\hat{n}})^{\mathsf{H}} \}$.

The filtering vector $\mathbf{f}_{\mathrm{CV},n}$ for $x_{n}$ is utilized to compute the LMMSE-ISIC estimate $\hat{x}_{n}$ and the filtering bias $\mu_{n}$, which are respectively given by
\begin{shrinkeq}{-0.5ex}
\begin{align}
\hat{x}_{n} = \mathbf{f}_{\mathrm{CV},n}^{\mathsf{H}}\tilde{\mathbf{y}}^{\hat{n}},
\end{align}
\end{shrinkeq}
and
\begin{shrinkeq}{-0.5ex}
\begin{align}
\mu_{n} = \mathbf{f}_{\mathrm{CV},n}^{\mathsf{H}}\mathbf{h}_{n}.
\end{align}
\end{shrinkeq}
Then, the residual interference-plus-noise variance can be written as $\eta_{n}^{2} = \mu_{n}(1 - \mu_{n})$. Based on $\hat{x}_{n}$, $\mu_{n}$, and $\eta_{n}^{2}$, the complex Gaussian approximation for each $x \in \mathcal{X}$ is given by $\psi_{n}(x) = \mathsf{exp}(\frac{-|\hat{x}_{n} - \mu_{n}x|^{2}}{\eta_{n}^{2}})$, which is then utilized to obtain the conditional probability $P_{n}(x) \triangleq P(x_{n} = x|\hat{x}_{n}) = \frac{\psi_{n}(x)}{\sum_{x' \in \mathcal{X}}\psi_{n}(x')}$ for each $x \in \mathcal{X}$. Finally, by leveraging $P_{n}(x)$'s for all $x \in \mathcal{X}$, the soft decision and residual interference variance of $x_{n}$ can be respectively given by $\bar{x}_{n} = \sum_{x \in \mathcal{X}}xP_{n}(x)$ and $v_{n} = \sum_{x \in \mathcal{X}}|x - \bar{x}_{n}|^{2}P_{n}(x)$. In any procedure for $x_{n}$ except the procedure for $x_{1}$ in the first iteration, $\bar{x}_{\mathsf{o}(n - 1)}$ and $v_{\mathsf{o}(n - 1)}$ are calculated in the previous procedure for $x_{\mathsf{o}(n - 1)}$ by the above two formulas, respectively. Note that since neither $\bar{x}_{n}$ nor $v_{n}$ is available before the first iteration, it is necessary to set the initial $\bar{x}_{n} = 0$ and $v_{n} = 1$ for $n \in \mathcal{N}$.

The procedure for $x_{n}$ has been mentioned above, while each LMMSE-ISIC iteration includes $N$ procedures for $ \{ x_{1}, \cdots, x_{N} \}$. After repeating the LMMSE-ISIC iteration until a stopping criterion (e.g., maximum number of iterations) is satisfied, $P_{n}(x)$'s (for all $x \in \mathcal{X}$) updated in the last iteration are utilized to determine the hard decision symbol $\check{x}_{n} = \mathsf{arg}\max_{x \in \mathcal{X}}P_{n}(x)$ for $n \in \mathcal{N}$.

\subsubsection{Low-Complexity Scheme}

To avoid the high computational complexity of the conventional LMMSE-ISIC scheme resulting from frequently computing matrix inverse, the affine MMSE (AMMSE) concept [33] was recently utilized by [9] to develop the low-complexity LMMSE-ISIC scheme, which replaces the SI-cancelled receive vector $\tilde{\mathbf{y}}^{\hat{n}}$ given in (12) and the filtering vector $\mathbf{f}_{\mathrm{CV},n}$ given in (13) with
\begin{shrinkeq}{-0.5ex}
\begin{align}
\tilde{\mathbf{y}} = \mathbf{y} - \sum_{i=1}^{N}\mathbf{h}_{i}\bar{x}_{i} = \mathbf{y} - \mathbf{H}\bar{\mathbf{x}},
\end{align}
\end{shrinkeq}
and
\begin{shrinkeq}{-0.5ex}
\begin{align}
\mathbf{f}_{\mathrm{PR},n} = (\mathbf{H}\mathbf{V}\mathbf{H}^{\mathsf{H}} + \sigma^{2}\mathbf{I}_{M})^{-1}\mathbf{h}_{n} = \mathbf{D}\mathbf{h}_{n},
\end{align}
\end{shrinkeq}
respectively, where $\mathbf{D} \triangleq (\mathbf{H}\mathbf{V}\mathbf{H}^{\mathsf{H}} + \sigma^{2}\mathbf{I}_{M})^{-1}$ represents the covariance matrix for $\tilde{\mathbf{y}}$, i.e., $\mathbf{D} = \mathbb{E} \{ \tilde{\mathbf{y}}\tilde{\mathbf{y}}^{\mathsf{H}} \}$.

Define the AMMSE filtering matrix as [24]
\begin{shrinkeq}{-0.5ex}
\begin{align}
\mathbf{F} \triangleq \mathbf{H}^{\mathsf{H}}\mathbf{D} = \mathbf{H}^{\mathsf{H}}(\mathbf{H}\mathbf{V}\mathbf{H}^{\mathsf{H}} + \sigma^{2}\mathbf{I}_{M})^{-1},
\end{align}
\end{shrinkeq}
and we have $\mathbf{f}_{\mathrm{PR},n} = [\mathbf{F}]_{n,:}^{\mathsf{H}}$. Then, according to the push-through identity $\mathbf{A}(\mathbf{B}\mathbf{A} + \mathbf{I}_{M})^{-1} = (\mathbf{A}\mathbf{B} + \mathbf{I}_{N})^{-1}\mathbf{A}$ [34], (18) can be recasted as
\begin{shrinkeq}{-0.5ex}
\begin{align}
\mathbf{F} = (\mathbf{H}^{\mathsf{H}}\mathbf{H}\mathbf{V} + \sigma^{2}\mathbf{I}_{N})^{-1}\mathbf{H}^{\mathsf{H}} = \mathbf{G}\mathbf{H}^{\mathsf{H}},
\end{align}
\end{shrinkeq}
where $\mathbf{G} \triangleq (\mathbf{H}^{\mathsf{H}}\mathbf{H}\mathbf{V} + \sigma^{2}\mathbf{I}_{N})^{-1}$, for simplifying the calculation due to usually $M \geq N$ as shown in Sec. II-A.

According to [24], the LMMSE-ISIC estimate $\hat{x}_{n}$ and the filtering bias $\mu_{n}$ are given by
\begin{shrinkeq}{-0.5ex}
\begin{align}
\hat{x}_{n} = \beta_{n}\mathbf{f}_{\mathrm{PR},n}^{\mathsf{H}}\tilde{\mathbf{y}} + \alpha_{n}\beta_{n}\hat{x}_{n},
\end{align}
\end{shrinkeq}
and
\begin{shrinkeq}{-0.5ex}
\begin{align}
\mu_{n} = \alpha_{n}\beta_{n},
\end{align}
\end{shrinkeq}
respectively, where $\alpha_{n} \triangleq \mathbf{f}_{\mathrm{PR},n}^{\mathsf{H}}\mathbf{h}_{n}$, $\beta_{n} \triangleq ((1 - v_{n})\alpha_{n} + 1)^{-1}$, and $\mathbf{f}_{\mathrm{PR},n}$ can be efficiently obtained by $\mathbf{f}_{\mathrm{PR},n} = \mathbf{H}[\mathbf{G}]_{n,:}^{\mathsf{H}}$. Meanwhile, to avoid directly computing matrix inverse, $\mathbf{G}$ can be updated by
\begin{shrinkeq}{-0.5ex}
\begin{align}
\mathbf{G} = \mathbf{G}^{'} - \frac{\mathbf{z}_{\mathsf{o}(n - 1)}[\mathbf{G}^{'}]_{\mathsf{o}(n - 1),:}}{[\mathbf{z}_{\mathsf{o}(n - 1)}]_{\mathsf{o}(n - 1)} + 1},
\end{align}
\end{shrinkeq}
where $\mathbf{G}^{'} \triangleq (\mathbf{H}^{\mathsf{H}}\mathbf{H}\mathbf{V}^{\mathsf{o}(n - 2)} + \sigma^{2}\mathbf{I}_{N})^{-1}$ and $\mathbf{z}_{\mathsf{o}(n - 1)} \triangleq (v_{\mathsf{o}(n - 1)} - v_{\mathsf{o}(n - 1)}^{\mathsf{o}(n - 2)})\mathbf{G}^{'}[\mathbf{H}^{\mathsf{H}}\mathbf{H}]_{\mathsf{o}(n - 1),:}^{\mathsf{H}}$.

The above necessary preliminaries will be utilized to elaborate our proposed relationship between LMMSE-ISIC and HD-OSIC (see Sec. III) and our proposed resursive scheme for LMMSE-ISIC (see Sec. IV).

\section{Proposed Relationship between LMMSE-ISIC and HD-OSIC}

In this section, we present a reordered description for the LMMSE-ISIC detection process based on the equivalent channel matrix, which is then compared with the HD-OSIC counterpart, to draw an essential conclusion that the former is the extension of the latter.

\vspace{-2mm}
\subsection{An Equivalent Channel Matrix Based Reordered Description for LMMSE-ISIC}

As shown in Sec. II-C, in the conventional description of the LMMSE-ISIC detection process for $x_{n}$ ($n \in \mathcal{N}$), $\bar{\mathbf{x}}^{\mathsf{o}(n - 2)}$ and $\mathbf{V}^{\mathsf{o}(n - 2)}$ are updated to $\bar{\mathbf{x}}^{\mathsf{o}(n - 1)}$ and $\mathbf{V}^{\mathsf{o}(n - 1)}$, respectively, by modifying the $\mathsf{o}(n - 1)$-th entry and diagonal entry. Moreover, it can be seen that $\bar{\mathbf{x}}^{\hat{n}}$ and $\mathbf{V}^{\hat{n}}$ can be obtained from $\bar{\mathbf{x}}^{\mathsf{o}(n - 1)}$ and $\mathbf{V}^{\mathsf{o}(n - 1)}$, respectively, by modifying the $n$-th entry and diagonal entry. Based on the above observations, we simplify the above procedure by proposing a reordered one. To this end, we utilize $\bar{x}_{n}^{n}$ and $v_{n}^{n}$ to immediately update $\bar{\mathbf{x}}$ and $\mathbf{V}$, respectively, and obtain $\bar{\mathbf{x}}^{n}$ and $\mathbf{V}^{n}$. Then, it can be verified that both $\bar{\mathbf{x}}^{\hat{n}}$ and $\bar{\mathbf{x}}^{n}$ can be obtained from $\bar{\mathbf{x}}^{\mathsf{o}(n - 1)}$ by modifying only the $n$-th entries, while both $\mathbf{V}^{\hat{n}}$ and $\mathbf{V}^{n}$ can be obtained from $\mathbf{V}^{\mathsf{o}(n - 1)}$ by modifying only the $n$-th diagonal entries. Specifically, $\bar{\mathbf{x}}^{n}$ and $\bar{\mathbf{x}}^{\hat{n}}$ can be obtained by replacing the $n$-th entry $\bar{x}_{n}^{\mathsf{o}(n - 1)}$ in $\bar{\mathbf{x}}^{\mathsf{o}(n - 1)}$ with $\bar{x}_{n}^{n}$ and
\begin{shrinkeq}{-0.5ex}
\begin{align}
\bar{x}_{n}^{\hat{n}} = 0,
\end{align}
\end{shrinkeq}
respectively, while $\mathbf{V}^{n}$ and $\mathbf{V}^{\hat{n}}$ can be obtained by replacing the $n$-th diagonal entry $v_{n}^{\mathsf{o}(n - 1)}$ in $\mathbf{V}^{\mathsf{o}(n - 1)}$ with $v_{n}^{n}$ and
\begin{shrinkeq}{-0.5ex}
\begin{align}
v_{n}^{\hat{n}} = 1,
\end{align}
\end{shrinkeq}
respectively. Therefore, we can conclude that $\bar{\mathbf{x}}^{\hat{n}}$ is just equal to $\bar{\mathbf{x}}^{n}$ with
\begin{shrinkeq}{-0.5ex}
\begin{align}
\bar{x}_{n}^{n} = 0,
\end{align}
\end{shrinkeq}
while $\mathbf{V}^{\hat{n}}$ is just equal to $\mathbf{V}^{n}$ with
\begin{shrinkeq}{-0.5ex}
\begin{align}
v_{n}^{n} = 1.
\end{align}
\end{shrinkeq}
In other words, we can regard the update (from $\bar{\mathbf{x}}^{\mathsf{o}(n - 1)}$ and $\mathbf{V}^{\mathsf{o}(n - 1)}$) to $\bar{\mathbf{x}}^{\hat{n}}$ and $\mathbf{V}^{\hat{n}}$ as a special case of that to $\bar{\mathbf{x}}^{n}$ and $\mathbf{V}^{n}$ with $\bar{x}_{n}^{n} = 0$ and $v_{n}^{n} = 1$.

For the convenience of description, we denote $\mathbf{V}^{\mathsf{o}(n - 1)}$, $\mathbf{V}^{\hat{n}}$ and $\mathbf{V}^{n}$ uniformly as $\mathbf{V}^{l}$, and denote $\bar{\mathbf{x}}^{\mathsf{o}(n - 1)}$, $\bar{\mathbf{x}}^{\hat{n}}$ and $\bar{\mathbf{x}}^{n}$ uniformly as $\bar{\mathbf{x}}^{l}$, where $l \in \{ \mathsf{o}(n - 1), \hat{n}, n \}$$^{\text{2}}$\footnote{$^{\text{2}}$If $\hat{n}$ needs to take a value, we can simply select $\hat{n} = n - \frac{1}{2}$ to distingush $\hat{n}$ from $\mathsf{o}(n - 1)$ and $n$.}. Similarly, we express the conventional SI-cancelled receive vector (12) and the AMMSE SI-cancelled receive vector (16) uniformly as $\tilde{\mathbf{y}}^{l}$, which is given by
\begin{shrinkeq}{-0.5ex}
\begin{align}
\tilde{\mathbf{y}}^{l} = \mathbf{y} - \mathbf{H}\bar{\mathbf{x}}^{l}, \; l \in \{ \mathsf{o}(n - 1), \hat{n}, n \}.
\end{align}
\end{shrinkeq}
On the other hand, the filtering vector $\mathbf{f}_{\mathrm{CV},n}$ given in (13) can be rewritten as
\begin{shrinkeq}{-0.5ex}
\begin{align}
\mathbf{f}_{\mathrm{CV},n} = [\mathbf{F}^{\hat{n}}]_{n,:}^{\mathsf{H}},
\end{align}
\end{shrinkeq}
where the filtering matrix $\mathbf{F}^{\hat{n}}$ is given as follows
\begin{shrinkeq}{-0.5ex}
\begin{align}
\mathbf{F}^{\hat{n}} = \mathbf{H}^{\mathsf{H}}\mathbf{D}^{\hat{n}} = \mathbf{H}^{\mathsf{H}}(\mathbf{H}\mathbf{V}^{\hat{n}}\mathbf{H}^{\mathsf{H}} + \sigma^{2}\mathbf{I}_{M})^{-1}.
\end{align}
\end{shrinkeq}
Then, we express the above conventional filtering matrix (29) and the AMMSE filtering matrix (18) uniformly as $\mathbf{F}^{l}$ ($l \in \{ \mathsf{o}(n - 1), \hat{n}, n \}$), which reads
\begin{shrinkeq}{-0.5ex}
\begin{align}
\mathbf{F}^{l} = \mathbf{H}^{\mathsf{H}}\mathbf{D}^{l} = \mathbf{H}^{\mathsf{H}}(\mathbf{H}\mathbf{V}^{l}\mathbf{H}^{\mathsf{H}} + \sigma^{2}\mathbf{I}_{M})^{-1}.
\end{align}
\end{shrinkeq}

To reduce the computational complexity, we first rewrite the above $\mathbf{F}^{l}$ as
\begin{shrinkeq}{-0.5ex}
\begin{align}
\mathbf{F}^{l} = (\mathbf{V}^{l})^{-\frac{1}{2}}(\tilde{\mathbf{H}}^{l})^{\mathsf{H}}(\tilde{\mathbf{H}}^{l}(\tilde{\mathbf{H}}^{l})^{\mathsf{H}} + \sigma^{2}\mathbf{I}_{M})^{-1},
\end{align}
\end{shrinkeq}
where the uniform \emph{equivalent channel matrix} $\tilde{\mathbf{H}}^{l}$ is defined as
\begin{shrinkeq}{-0.5ex}
\begin{align}
\tilde{\mathbf{H}}^{l} \triangleq \mathbf{H}(\mathbf{V}^{l})^{\frac{1}{2}}.
\end{align}
\end{shrinkeq}
Then, by applying the push-through identity $\mathbf{A}(\mathbf{B}\mathbf{A} + \mathbf{I}_{M})^{-1} = (\mathbf{A}\mathbf{B} + \mathbf{I}_{N})^{-1}\mathbf{A}$ [34], (31) can be further transformed as
\begin{shrinkeq}{-0.5ex}
\begin{align}
\mathbf{F}^{l} = (\mathbf{V}^{l})^{-\frac{1}{2}}((\tilde{\mathbf{H}}^{l})^{\mathsf{H}}\tilde{\mathbf{H}}^{l} + \sigma^{2}\mathbf{I}_{N})^{-1}(\tilde{\mathbf{H}}^{l})^{\mathsf{H}},
\end{align}
\end{shrinkeq}
which replaces the $M \times M$ matrix inverse in (31) with the usually smaller $N \times N$ inverse. Accordingly, with $l = \hat{n}$, the filtering matrix $\mathbf{F}^{\hat{n}}$ can be thus obtained, and based on (28), the filtering vector $\mathbf{f}_{\mathrm{CV},n}$ can be expressed as
\begin{shrinkeq}{-0.5ex}
\begin{align}
\mathbf{f}_{\mathrm{CV},n} &= [(\mathbf{V}^{\hat{n}})^{-\frac{1}{2}}((\tilde{\mathbf{H}}^{\hat{n}})^{\mathsf{H}}\tilde{\mathbf{H}}^{\hat{n}} + \sigma^{2}\mathbf{I}_{N})^{-1}(\tilde{\mathbf{H}}^{\hat{n}})^{\mathsf{H}}]_{n,:}^{\mathsf{H}} \notag\\
&= \frac{1}{\sqrt{v_{n}^{\hat{n}}}}[((\tilde{\mathbf{H}}^{\hat{n}})^{\mathsf{H}}\tilde{\mathbf{H}}^{\hat{n}} + \sigma^{2}\mathbf{I}_{N})^{-1}(\tilde{\mathbf{H}}^{\hat{n}})^{\mathsf{H}}]_{n,:}^{\mathsf{H}}.
\end{align}
\end{shrinkeq}
Finally, according to $\hat{x}_{n} = \mathbf{f}_{\mathrm{CV},n}^{\mathsf{H}}\tilde{\mathbf{y}}^{\hat{n}}$ given in (14), the LMMSE-ISIC estimate of $x_{n}$ is given as follows
\begin{shrinkeq}{-0.5ex}
\begin{align}
\hat{x}_{n} = \frac{1}{\sqrt{v_{n}^{\hat{n}}}}[((\tilde{\mathbf{H}}^{\hat{n}})^{\mathsf{H}}\tilde{\mathbf{H}}^{\hat{n}} + \sigma^{2}\mathbf{I}_{N})^{-1}(\tilde{\mathbf{H}}^{\hat{n}})^{\mathsf{H}}]_{n,:}\tilde{\mathbf{y}}^{\hat{n}}.
\end{align}
\end{shrinkeq}

\subsection{LMMSE-ISIC: Extension of HD-OSIC}

Recalling $\hat{x}_{n} = [(\mathbf{H}_{n}^{\mathsf{H}}\mathbf{H}_{n} + \sigma^{2}\mathbf{I}_{n})^{-1}\mathbf{H}_{n}^{\mathsf{H}}]_{n,:}\mathbf{y}^{(n)}$ given in (2) for computing the estimate under HD-OSIC schemes, it can be observed that (35) for LMMSE-ISIC is mathematically identical to (2) for HD-OSIC, while the equivalent channel matrix $\tilde{\mathbf{H}}^{\hat{n}}$ and the SI-cancelled receive vector $\tilde{\mathbf{y}}^{\hat{n}}$ within (35) are mathematically identical to the channel matrix $\mathbf{H}_{n}$ and the HI-cancelled receive vector $\mathbf{y}^{(n)}$ within (2), respectively. In this subsection, we will elaborate that the update of the uniform SI-cancelled received vector $\tilde{\mathbf{y}}^{l}$ for LMMSE-ISIC is mathematically identical to the update of the HI-cancelled receive vector $\mathbf{y}^{(n)}$ for HD-OSIC, while the update of the uniform equivalent channel matrix $\tilde{\mathbf{H}}^{l}$ for LMMSE-ISIC can be regarded as the extension of the update of the channel matrix $\mathbf{H}_{n}$ for HD-OSIC. Accordingly, we achieve the final conclusion that the reordered description of the LMMSE-ISIC detection process can be regarded as the extension of the HD-OSIC detection process.

As aforementioned, in the procedure for $x_{n}$, the LMMSE-ISIC detector needs to update $\tilde{\mathbf{H}}^{\mathsf{o}(n - 1)}$ and $\tilde{\mathbf{y}}^{\mathsf{o}(n - 1)}$ to $\tilde{\mathbf{H}}^{\hat{n}}$ and $\tilde{\mathbf{y}}^{\hat{n}}$, respectively, for computing $\hat{x}_{n}$, and also needs to update them to $\tilde{\mathbf{H}}^{n}$ and $\tilde{\mathbf{y}}^{n}$, respectively, for the next procedure. Note that $\tilde{\mathbf{H}}^{l}$ and $\tilde{\mathbf{y}}^{l}$ (for $l \in \{ \mathsf{o}(n - 1), \hat{n}, n \}$) are related to $\mathbf{V}^{l}$ and $\bar{\mathbf{x}}^{l}$, respectively, as shown in (32) and (27). Besides, it has been verified that $\mathbf{V}^{\hat{n}}$ is equal to $\mathbf{V}^{n}$ when $v_{n}^{n} = 1$, and $\bar{\mathbf{x}}^{\hat{n}}$ is equal to $\bar{\mathbf{x}}^{n}$ when $\bar{x}_{n}^{n} = 0$. Accordingly, we can conclude that $\tilde{\mathbf{H}}^{\hat{n}}$ is equal to $\tilde{\mathbf{H}}^{n}$ if $v_{n}^{n}$ in $\mathbf{V}^{n}$ satisfies $v_{n}^{n} = 1$, while $\tilde{\mathbf{y}}^{\hat{n}}$ is equal to $\tilde{\mathbf{y}}^{n}$ if $\bar{x}_{n}^{n}$ in $\bar{\mathbf{x}}^{n}$ satisfies $\bar{x}_{n}^{n} = 0$. Hence in the sequel, we only discuss the procedures for updating $\tilde{\mathbf{H}}^{\mathsf{o}(n - 1)}$ and $\tilde{\mathbf{y}}^{\mathsf{o}(n - 1)}$ to $\tilde{\mathbf{H}}^{n}$ and $\tilde{\mathbf{y}}^{n}$, respectively. For brevity, in the rest of this paper, we will always use $\bar{\mathbf{x}}^{'}$, $\mathbf{V}^{'}$, $\tilde{\mathbf{H}}^{'}$, and $\tilde{\mathbf{y}}^{'}$ as the shorthand expressions for the above mentioned $\bar{\mathbf{x}}^{\mathsf{o}(n - 1)}$, $\mathbf{V}^{\mathsf{o}(n - 1)}$, $\tilde{\mathbf{H}}^{\mathsf{o}(n - 1)}$, and $\tilde{\mathbf{y}}^{\mathsf{o}(n - 1)}$, respectively. Besides, in the rest of this subsection, we only consider the procedure for $x_{N}$ (i.e., the case of $n = N$), and the procedures for $x_{n}$ ($n \in \mathcal{N}$) other than $n = N$ can be similarly acquired, which is omitted for brevity.

Recall the soft decision vector $\bar{\mathbf{x}} = [\bar{x}_{1}, \cdots, \bar{x}_{N}]^{\mathsf{T}}$ and the residual interference variance matrix $\mathbf{V} = \mathsf{Diag}([v_{1}, \cdots, v_{N}]^{\mathsf{T}})$. By replacing $\bar{x}_{N}^{'}$ in $\bar{\mathbf{x}}^{'}$ and $v_{N}^{'}$ in $\mathbf{V}^{'}$ with $\bar{x}_{N}^{N}$ and $v_{N}^{N}$, respectively, $\bar{\mathbf{x}}^{'}$ and $\mathbf{V}^{'}$ are updated to $\bar{\mathbf{x}}^{N}$ and $\mathbf{V}^{N}$, respectively, i.e.,
\begin{shrinkeq}{-0.5ex}
\begin{align}
\bar{\mathbf{x}}^{N} &= [\bar{\mathbf{x}}_{N - 1}^{'\mathsf{T}}, \bar{x}_{N}^{N}]^{\mathsf{T}}, \\
\mathbf{V}^{N} &= \begin{bmatrix}
\mathbf{V}_{N - 1}^{'} & \bm{0}_{N - 1}\\
\bm{0}_{N - 1}^{\mathsf{T}} & v_{N}^{N}\\
\end{bmatrix},
\end{align}
\end{shrinkeq}
where $\bar{\mathbf{x}}_{N - 1}^{'}$ collects the first $(N - 1)$ entries of $\bar{\mathbf{x}}^{'}$, and $\mathbf{V}_{N - 1}^{'}$ represents the $(N - 1) \times (N - 1)$ leading principal submatrix of $\mathbf{V}^{'}$. The above (36) and (37) will be utilized to update the uniform SI-cancelled receive vector $\tilde{\mathbf{y}}^{'}$ and the uniform equivalent channel matrix $\tilde{\mathbf{H}}^{'}$ to $\tilde{\mathbf{y}}^{N}$ and $\tilde{\mathbf{H}}^{N}$, respectively, which is provided by the following theorem.
\begin{thm}
The uniform SI-cancelled receive vector $\tilde{\mathbf{y}}^{'}$ and the uniform equivalent channel matrix $\tilde{\mathbf{H}}^{'}$ can be respectively updated to $\tilde{\mathbf{y}}^{N}$ and $\tilde{\mathbf{H}}^{N}$ by
\begin{shrinkeq}{-0.5ex}
\begin{align}
\tilde{\mathbf{y}}^{N} &= \tilde{\mathbf{y}}^{'} - \frac{\bar{x}_{N}^{N} - \bar{x}_{N}^{'}}{\sqrt{v_{N}^{'}}}\tilde{\mathbf{h}}_{N}^{'}, \\
\tilde{\mathbf{H}}^{N} &= \bigg[\tilde{\mathbf{H}}_{N - 1}^{'}, \sqrt{\frac{v_{N}^{N}}{v_{N}^{'}}}\tilde{\mathbf{h}}_{N}^{'} \bigg],
\end{align}
\end{shrinkeq}
where $\tilde{\mathbf{H}}_{N - 1}^{'}$ and $\tilde{\mathbf{h}}_{N}^{'}$ denote the first $(N - 1)$ columns and the $N$-th column of $\tilde{\mathbf{H}}^{'}$, respectively.
\end{thm}
\begin{proof}
Taking (38) into account first, by substituting (36) into $\tilde{\mathbf{y}}^{l} = \mathbf{y} - \mathbf{H}\bar{\mathbf{x}}^{l}$ given in (27) with $l = N$, we have
\begin{shrinkeq}{-0.5ex}
\begin{align}
\tilde{\mathbf{y}}^{N} &= \mathbf{y} - \mathbf{H}[\bar{\mathbf{x}}_{N - 1}^{'\mathsf{T}}, \bar{x}_{N}^{N}]^{\mathsf{T}} = \mathbf{y} - \mathbf{H}_{N - 1}\bar{\mathbf{x}}_{N - 1}^{'} - \mathbf{h}_{N}\bar{x}_{N}^{N} \notag\\
&= \mathbf{y} - \mathbf{H}\bar{\mathbf{x}}^{'} + \mathbf{h}_{N}\bar{x}_{N}^{'} - \mathbf{h}_{N}\bar{x}_{N}^{N} \notag\\
&= \tilde{\mathbf{y}}^{'} + \mathbf{h}_{N}\bar{x}_{N}^{'} - \mathbf{h}_{N}\bar{x}_{N}^{N},
\end{align}
\end{shrinkeq}
where the last equality follows (27) with $l = \mathsf{o}(N - 1)$. Besides, according to $\tilde{\mathbf{H}}^{l} = \mathbf{H}(\mathbf{V}^{l})^{\frac{1}{2}}$ given in (32), the $n$-th column of $\mathbf{H}$, i.e., $\mathbf{h}_{n}$, can be given by
\begin{shrinkeq}{-0.5ex}
\begin{align}
\mathbf{h}_{n} = \frac{\tilde{\mathbf{h}}_{n}^{l}}{\sqrt{v_{n}^{l}}}.
\end{align}
\end{shrinkeq}
Therefore, by substituting (41) with $l = \mathsf{o}(N - 1)$ into (40), we obtain (38).

Next, we proceed to derive (39). By substituting (37) into (32) with $l = N$, we have
\begin{shrinkeq}{-0.5ex}
\begin{align}
\tilde{\mathbf{H}}^{N} &= \mathbf{H}\begin{bmatrix}
\mathbf{V}_{N - 1}^{'} & \bm{0}_{N - 1}\\
\bm{0}_{N - 1}^{\mathsf{T}} & v_{N}^{N}\\
\end{bmatrix}^{\frac{1}{2}} = [\mathbf{H}_{N - 1}\mathbf{V}_{N - 1}^{'\frac{1}{2}}, \sqrt{v_{N}^{N}}\mathbf{h}_{N}] \notag\\
&= [\tilde{\mathbf{H}}_{N - 1}^{'}, \sqrt{v_{N}^{N}}\mathbf{h}_{N}],
\end{align}
\end{shrinkeq}
where the last equality follows (32) with $l = \mathsf{o}(N - 1)$. Again, by substituting (41) into (42), we obtain (39).
\end{proof}

According to \textbf{Theorem 1} and $\mathbf{y}^{(n - 1)} = \mathbf{y}^{(n)} - \check{x}_{n}\mathbf{h}_{n}$ given in (3), it can be observed that the update of the interference cancelled receive vector given in (38) for LMMSE-ISIC is mathematically identical to (3) for HD-OSIC, while $\frac{\bar{x}_{N}^{N} - \bar{x}_{N}^{'}}{\sqrt{v_{N}^{'}}}$ in (38) is equivalent to $\check{x}_{n}$ in (3). On the other hand, it can be observed from (39) that the last column of $\tilde{\mathbf{H}}^{'}$ is multiplied by a scalar to update $\tilde{\mathbf{H}}^{'}$ to $\tilde{\mathbf{H}}^{N}$ for LMMSE-ISIC, while the last column of the permuted $\mathbf{H}$ is removed to deflate $\mathbf{H}_{N} = [\mathbf{h}_{1}, \cdots, \mathbf{h}_{N}]$ to $\mathbf{H}_{N - 1} = [\mathbf{h}_{1}, \cdots, \mathbf{h}_{N - 1}]$ for HD-OSIC. Obviously, $\tilde{\mathbf{H}}^{'}$ can be deflated to $\tilde{\mathbf{H}}_{N - 1}^{'}$ by (39) if $v_{N}^{N} = 0$, in other words, the update of $\mathbf{H}_{N}$ to $\mathbf{H}_{N - 1}$ for HD-OSIC can be regarded as a special case of the update of $\tilde{\mathbf{H}}^{'}$ to $\tilde{\mathbf{H}}^{N}$ for LMMSE-ISIC. Accordingly, the update of the uniform equivalent channel matrix given in (39) for LMMSE-ISIC can be regarded as the extension of the update of the channel matrix for HD-OSIC. Therefore, we can finally conclude that the LMMSE-ISIC detection process constituted by (35), (38) and (39) can be regarded as the extension of the HD-OSIC detection process constituted by (2), (3) and $\mathbf{H}_{n} = [\mathbf{h}_{1}, \cdots, \mathbf{h}_{n}]$.

\vspace{-2mm}
\section{Proposed Recursive Scheme for LMMSE-ISIC}

In the last section, we obtain an important insight that the LMMSE-ISIC detection process is the extension of the HD-OSIC counterpart. Based on this conclusion, in this section, we propose a recursive scheme for LMMSE-ISIC by extending that for HD-OSIC.

\vspace{-2mm}
\subsection{Algorithm Design}

Our proposed recursive scheme for LMMSE-ISIC is based on the following Hermitian inverse matrix
\begin{shrinkeq}{-0.5ex}
\begin{align}
\mathbf{Q}_{n}^{l} = ((\tilde{\mathbf{H}}_{n}^{l})^{\mathsf{H}}\tilde{\mathbf{H}}_{n}^{l} + \sigma^{2}\mathbf{I}_{n})^{-1}, \; l \in \{ \mathsf{o}(n - 1), \hat{n}, n \},
\end{align}
\end{shrinkeq}
where $\tilde{\mathbf{H}}_{n}^{l} \triangleq [\tilde{\mathbf{h}}_{1}^{l}, \cdots, \tilde{\mathbf{h}}_{n}^{l}]$ collects the first $n$ columns of $\tilde{\mathbf{H}}^{l}$, since the recursive HD-OSIC scheme is based on $\mathbf{Q}_{n}$ related to $\mathbf{H}_{n}$ (see Sec. II-B). Then, the computation of the LMMSE-ISIC estimate $\hat{x}_{n}$ is given by the following proposition.
\begin{prop}
The LMMSE-ISIC estimate $\hat{x}_{n}$ can be obtained by
\begin{shrinkeq}{-0.5ex}
\begin{align}
\hat{x}_{n} = [\tilde{\mathbf{t}}^{\hat{n}}]_{n},
\end{align}
\end{shrinkeq}
via applying the uniform symbol estimate vector $\tilde{\mathbf{t}}^{l}$ related to $\mathbf{Q}_{N}^{l}$ with $l = \hat{n}$, which is given by
\begin{shrinkeq}{-0.5ex}
\begin{align}
\tilde{\mathbf{t}}^{l} = (\mathbf{V}^{l})^{-\frac{1}{2}}\mathbf{Q}_{N}^{l}(\tilde{\mathbf{H}}^{l})^{\mathsf{H}}\tilde{\mathbf{y}}^{l}.
\end{align}
\end{shrinkeq}
\end{prop}
\begin{proof}
Actually, $\tilde{\mathbf{t}}^{l}$ is defined as
\begin{shrinkeq}{-0.5ex}
\begin{align}
\tilde{\mathbf{t}}^{l} \triangleq \mathbf{F}^{l}\tilde{\mathbf{y}}^{l},
\end{align}
\end{shrinkeq}
where the uniform filtering matrix $\mathbf{F}^{l}$ and the uniform SI-cancelled receive vector $\tilde{\mathbf{y}}^{l}$ have been given in (33) and (27), respectively. Then, by substituting (43) with $n = N$ into (33), we obtain
\begin{shrinkeq}{-0.5ex}
\begin{align}
\mathbf{F}^{l} = (\mathbf{V}^{l})^{-\frac{1}{2}}\mathbf{Q}_{N}^{l}(\tilde{\mathbf{H}}^{l})^{\mathsf{H}},
\end{align}
\end{shrinkeq}
and (45) can be readily obtained by substituting (47) into (46). Finally, by replacing $l$ in (46) with $\hat{n}$ and substituting $\mathbf{f}_{\mathrm{CV},n} = [\mathbf{F}^{\hat{n}}]_{n,:}^{\mathsf{H}}$ shown in (28) into $\hat{x}_{n} = \mathbf{f}_{\mathrm{CV},n}^{\mathsf{H}}\tilde{\mathbf{y}}^{\hat{n}}$ shown in (14), we have
\begin{shrinkeq}{-0.5ex}
\begin{align}
\hat{x}_{n} = [\mathbf{F}^{\hat{n}}\tilde{\mathbf{y}}^{\hat{n}}]_{n} = [\tilde{\mathbf{t}}^{\hat{n}}]_{n},
\end{align}
\end{shrinkeq}
and thus (44) can be obtained.
\end{proof}

\emph{Remark 1: It can be observed from \textbf{Proposition 1} that (44) and (45) are just similar to those for HD-OSIC (see Sec. II-B).}

Moreover, the filtering bias $\mu_{n}$ is computed by leveraging the following proposition.
\begin{prop}
Via utilizing the $n$-th diagonal entry of $\mathbf{Q}_{N}^{\hat{n}}$ (i.e., $\mathbf{Q}_{N}^{l}$ with $l = \hat{n}$), the computation of the filtering bias $\mu_{n}$ can be expressed as
\begin{shrinkeq}{-0.5ex}
\begin{align}
\mu_{n} = 1 - \sigma^{2}[\mathbf{Q}_{N}^{\hat{n}}]_{n,n}.
\end{align}
\end{shrinkeq}
\end{prop}
\begin{proof}
By substituting (43) and $v_{n}^{\hat{n}} = 1$ given in (24) into (34), we obtain
\begin{shrinkeq}{-0.5ex}
\begin{align}
\mathbf{f}_{\mathrm{CV},n}^{\mathsf{H}} = [\mathbf{Q}_{N}^{\hat{n}}(\tilde{\mathbf{H}}^{\hat{n}})^{\mathsf{H}}]_{n,:} = [\mathbf{Q}_{N}^{\hat{n}}]_{n,:}(\tilde{\mathbf{H}}^{\hat{n}})^{\mathsf{H}}.
\end{align}
\end{shrinkeq}
Then, by substituting $\mathbf{h}_{n} = \frac{\tilde{\mathbf{h}}_{n}^{l}}{\sqrt{v_{n}^{l}}}$ given in (41) with $l = \hat{n}$ and (50) into $\mu_{n} = \mathbf{f}_{\mathrm{CV},n}^{\mathsf{H}}\mathbf{h}_{n}$ given in (15), we have
\begin{shrinkeq}{-0.5ex}
\begin{align}
\mu_{n} = [\mathbf{Q}_{N}^{\hat{n}}]_{n,:}(\tilde{\mathbf{H}}^{\hat{n}})^{\mathsf{H}}\frac{\tilde{\mathbf{h}}_{n}^{\hat{n}}}{\sqrt{v_{n}^{\hat{n}}}}.
\end{align}
\end{shrinkeq}
Note that we can utilize (43) to express $\mathbf{Q}_{N}^{l}(\tilde{\mathbf{H}}^{l})^{\mathsf{H}}\tilde{\mathbf{H}}^{l}$ as
\begin{shrinkeq}{-0.5ex}
\begin{align}
\mathbf{Q}_{N}^{l}(\tilde{\mathbf{H}}^{l})^{\mathsf{H}}\tilde{\mathbf{H}}^{l} &= ((\tilde{\mathbf{H}}^{l})^{\mathsf{H}}\tilde{\mathbf{H}}^{l} + \sigma^{2}\mathbf{I}_{N})^{-1}(\tilde{\mathbf{H}}^{l})^{\mathsf{H}}\tilde{\mathbf{H}}^{l} \notag\\
&= \mathbf{I}_{N} - \sigma^{2}((\tilde{\mathbf{H}}^{l})^{\mathsf{H}}\tilde{\mathbf{H}}^{l} + \sigma^{2}\mathbf{I}_{N})^{-1} \notag\\
&= \mathbf{I}_{N} - \sigma^{2}\mathbf{Q}_{N}^{l},
\end{align}
\end{shrinkeq}
whose $n$-th ($n \in \mathcal{N}$) diagonal entry is given by
\begin{shrinkeq}{-0.5ex}
\begin{align}
[\mathbf{Q}_{N}^{l}(\tilde{\mathbf{H}}^{l})^{\mathsf{H}}\tilde{\mathbf{H}}^{l}]_{n,n} \!=\! [\mathbf{Q}_{N}^{l}]_{n,:}(\tilde{\mathbf{H}}^{l})^{\mathsf{H}}\tilde{\mathbf{h}}_{n}^{l} \!=\! 1 \!-\! \sigma^{2}[\mathbf{Q}_{N}^{l}]_{n,n}.
\end{align}
\end{shrinkeq}
Hence, by substituting (53) with $l = \hat{n}$ and $v_{n}^{\hat{n}} = 1$ given in (24) into (51), (49) can be achieved.
\end{proof}

As suggested by the above two propositions, the LMMSE-ISIC estimate and the filtering bias can be readily computed from the entries of $\tilde{\mathbf{t}}^{\hat{n}}$ and $\mathbf{Q}_{N}^{\hat{n}}$, respectively. Therefore, we only need to develop the algorithms to acquire $\tilde{\mathbf{t}}^{\hat{n}}$ and $\mathbf{Q}_{N}^{\hat{n}}$. Specifically, we first update $\mathbf{Q}_{N}^{\mathsf{o}(n - 1)}$ and $\tilde{\mathbf{t}}^{\mathsf{o}(n - 1)}$ to $\mathbf{Q}_{N}^{n}$ and $\tilde{\mathbf{t}}^{n}$, respectively. Then, we substitute $\bar{x}_{n}^{n} = 0$ given in (25) and $v_{n}^{n} = 1$ given in (26) into the above procedure to achieve the update from $\mathbf{Q}_{N}^{\mathsf{o}(n - 1)}$ and $\tilde{\mathbf{t}}^{\mathsf{o}(n - 1)}$ to $\mathbf{Q}_{N}^{\hat{n}}$ and $\tilde{\mathbf{t}}^{\hat{n}}$, respectively. Note that in the remainder of this paper, $\mathbf{Q}^{'}$ and $\tilde{\mathbf{t}}^{'}$ will always represent $\mathbf{Q}^{\mathsf{o}(n - 1)}$ and $\tilde{\mathbf{t}}^{\mathsf{o}(n - 1)}$, respectively.

Taking the case of $n = N$ into account first, we design the procedure for $x_{N}$. For one thing, the following proposition describes the method for updating the inverse matrix $\mathbf{Q}_{N}^{'}$ to $\mathbf{Q}_{N}^{N}$.
\begin{prop}
The inverse matrix $\mathbf{Q}_{N}^{'}$ can be directly updated to $\mathbf{Q}_{N}^{N}$ by
\begin{shrinkeq}{-0.5ex}
\begin{align}
\mathbf{Q}_{N}^{N} = \begin{bmatrix}
\ddot{\mathbf{Q}}_{N - 1}^{N} & \ddot{\mathbf{q}}_{N}^{N}\\
(\ddot{\mathbf{q}}_{N}^{N})^{\mathsf{H}} & \omega_{N}^{N}\\
\end{bmatrix}
\end{align}
\end{shrinkeq}
with
\begin{shrinkeq}{-0.5ex}
\begin{align}
\omega_{N}^{N} &= \frac{\omega_{N}^{'}v_{N}^{'}}{v_{N}^{N} + \sigma^{2}\omega_{N}^{'}(v_{N}^{'} - v_{N}^{N})}, \\
\ddot{\mathbf{q}}_{N}^{N} &= \frac{\omega_{N}^{N}}{\omega_{N}^{'}}\sqrt{\frac{v_{N}^{N}}{v_{N}^{'}}}\ddot{\mathbf{q}}_{N}^{'}, \\
\ddot{\mathbf{Q}}_{N - 1}^{N} &= \ddot{\mathbf{Q}}_{N - 1}^{'} + \frac{\omega_{N}^{N}v_{N}^{N} - \omega_{N}^{'}v_{N}^{'}}{\omega_{N}^{'2}v_{N}^{'}}\ddot{\mathbf{q}}_{N}^{'}\ddot{\mathbf{q}}_{N}^{'\mathsf{H}},
\end{align}
\end{shrinkeq}
where $\ddot{\mathbf{Q}}_{N - 1}^{'}$, $\ddot{\mathbf{q}}_{N}^{'}$, and $\omega_{N}^{'}$ are in $\mathbf{Q}_{N}^{'}$, as the structure in (54).
\end{prop}
\begin{proof}
See Appendix A.
\end{proof}

For another, the method for updating the uniform symbol estimate vector $\tilde{\mathbf{t}}^{'}$ to $\tilde{\mathbf{t}}^{N}$ is given by the following proposition.
\begin{prop}
The uniform symbol estimate vector $\tilde{\mathbf{t}}^{'}$ can be updated to $\tilde{\mathbf{t}}^{N}$ by
\begin{shrinkeq}{-0.5ex}
\begin{align}
\tilde{\mathbf{t}}_{N - 1}^{N} &= \tilde{\mathbf{t}}_{N - 1}^{'} + \sigma^{2}\sqrt{v_{N}^{'}}(\mathbf{V}_{N - 1}^{N})^{-\frac{1}{2}}\ddot{\mathbf{q}}_{N}^{'} \notag\\
& \times \frac{\bar{x}_{N}^{N} - \bar{x}_{N}^{'} + (v_{N}^{N} - v_{N}^{'})[\tilde{\mathbf{t}}^{'}]_{N}}{v_{N}^{N} - \sigma^{2}\omega_{N}^{'}(v_{N}^{N} - v_{N}^{'})}, \\
[\tilde{\mathbf{t}}^{N}]_{N} &= \frac{v_{N}^{'}[\tilde{\mathbf{t}}^{'}]_{N} + (\bar{x}_{N}^{N} - \bar{x}_{N}^{'})(\sigma^{2}\omega_{N}^{'} - 1)}{v_{N}^{N} - \sigma^{2}\omega_{N}^{'}(v_{N}^{N} - v_{N}^{'})},
\end{align}
\end{shrinkeq}
where $\tilde{\mathbf{t}}_{N - 1}^{N}$ denotes the first $(N - 1)$ entries of $\tilde{\mathbf{t}}^{N}$.
\end{prop}
\begin{proof}
See Appendix B.
\end{proof}

\emph{Remark 2: Recalling $v_{n} = \sum_{x \in \mathcal{X}}|x - \bar{x}_{n}|^{2}P_{n}(x)$ in Sec. II-C, it can be observed that when $v_{n} \rightarrow 0$, $\bar{x}_{n} \rightarrow x \in \mathcal{X}$, i.e., the soft decision $\bar{x}_{n}$ approaches the hard decision $\check{x}_{n}$. By substituting $v_{n}^{n} \rightarrow 0$ with $n = N$ into (56) and (57), we obtain}
\begin{shrinkeq}{-0.5ex}
\begin{align}
\ddot{\mathbf{q}}_{N}^{N} &= \bm{0}_{N - 1}, \\
\ddot{\mathbf{Q}}_{N - 1}^{N} &= \ddot{\mathbf{Q}}_{N - 1}^{'} - \frac{1}{\omega_{N}^{'}}\ddot{\mathbf{q}}_{N}^{'}\ddot{\mathbf{q}}_{N}^{'\mathsf{H}}.
\end{align}
\end{shrinkeq}
\emph{Meanwhile, by substituting $v_{n}^{n} \rightarrow 0$ with $n = N$ into (58), we obtain}
\begin{shrinkeq}{-0.5ex}
\begin{align}
\tilde{\mathbf{t}}_{N - 1}^{N} = \tilde{\mathbf{t}}_{N - 1}^{'} + \frac{\bar{x}_{N}^{N} - \bar{x}_{N}^{'} - v_{N}^{'}[\tilde{\mathbf{t}}^{'}]_{N}}{\omega_{N}^{'}\sqrt{v_{N}^{'}}}(\mathbf{V}_{N - 1}^{N})^{-\frac{1}{2}}\ddot{\mathbf{q}}_{N}^{'}.
\end{align}
\end{shrinkeq}
\emph{Obviously, (60) and (61) are equivalent to those for deflating $\mathbf{Q}_{n}$ to $\mathbf{Q}_{n - 1}$ in HD-OSIC, while (62) is similar to that for updating the symbol estimate vector in HD-OSIC (details can be found in Sec. II-B). Hence, the recursive HD-OSIC scheme can be regarded as the special case with $v_{n}^{n} \rightarrow 0$ of our proposed recursive LMMSE-ISIC scheme.}

Up to this point, the procedure for $x_{N}$, i.e., (54)-(59), has been developed. In the following, we directly present the procedure for $x_{n}$ ($n \in \mathcal{N}$), due to complicated derivations and space limitation. When applied to any procedure for $x_{n}$ ($n \in \mathcal{N}$), (54)-(57) will be extended to
\begin{shrinkeq}{-0.5ex}
\begin{align}
&[\mathbf{Q}_{N}^{n}]_{n,n} = \omega_{n}^{n} = \frac{\omega_{n}^{'}v_{n}^{'}}{v_{n}^{n} + \sigma^{2}\omega_{n}^{'}(v_{n}^{'} - v_{n}^{n})}, \\
&\mathbf{Q}_{N - 1}^{n[-n]} \!=\! \mathbf{Q}_{N - 1}^{'[-n]} \!+\! \frac{\omega_{n}^{n}v_{n}^{n} \!-\! \omega_{n}^{'}v_{n}^{'}}{(\omega_{n}^{'})^{2}v_{n}^{'}}[\mathbf{Q}_{N}^{'}]_{:,n}^{[-n]}([\mathbf{Q}_{N}^{'}]_{:,n}^{[-n]})^{\mathsf{H}}, \\
&[\mathbf{Q}_{N}^{n}]_{:,n}^{[-n]} = \frac{\omega_{n}^{n}}{\omega_{n}^{'}}\sqrt{\frac{v_{n}^{n}}{v_{n}^{'}}}[\mathbf{Q}_{N}^{'}]_{:,n}^{[-n]}, \\
&[\mathbf{Q}_{N}^{n}]_{n,:}^{[-n]} = ([\mathbf{Q}_{N}^{n}]_{:,n}^{[-n]})^{\mathsf{H}},
\end{align}
\end{shrinkeq}
where $(\cdot)^{'}$ represents $(\cdot)^{\mathsf{o}(n - 1)}$ for brevity, $\mathbf{Q}_{N - 1}^{n[-n]}$ represents the matrix $\mathbf{Q}_{N}^{n}$ with its $n$-th row and column removed, $[\mathbf{Q}_{N}^{n}]_{:,n}^{[-n]}$ represents the $n$-th column of $\mathbf{Q}_{N}^{n}$ with its $n$-th entry removed, and $[\mathbf{Q}_{N}^{n}]_{n,:}^{[-n]}$ represents the $n$-th row of $\mathbf{Q}_{N}^{n}$ with its $n$-th entry removed. $\mathbf{Q}_{N - 1}^{'[-n]}$ and $[\mathbf{Q}_{N}^{'}]_{:,n}^{[-n]}$ are similarly defined, while $\omega_{n}^{'} = [\mathbf{Q}_{N}^{'}]_{n,n}$. On the other hand, (58) and (59) can be respectively extended to
\begin{shrinkeq}{-0.5ex}
\begin{align}
\tilde{\mathbf{t}}_{N - 1}^{n[-n]} &= \tilde{\mathbf{t}}_{N - 1}^{'[-n]} + \sigma^{2}\sqrt{v_{n}^{'}}(\mathbf{V}_{N - 1}^{n[-n]})^{-\frac{1}{2}}[\mathbf{Q}_{N}^{'}]_{:,n}^{[-n]} \notag\\
& \times \frac{\bar{x}_{n}^{n} - \bar{x}_{n}^{'} + (v_{n}^{n} - v_{n}^{'})[\tilde{\mathbf{t}}^{'}]_{n}}{v_{n}^{n} - \sigma^{2}\omega_{n}^{'}(v_{n}^{n} - v_{n}^{'})}, \\
[\tilde{\mathbf{t}}^{n}]_{n} &= \frac{v_{n}^{'}[\tilde{\mathbf{t}}^{'}]_{n} + (\bar{x}_{n}^{n} - \bar{x}_{n}^{'})(\sigma^{2}\omega_{n}^{'} - 1)}{v_{n}^{n} - \sigma^{2}\omega_{n}^{'}(v_{n}^{n} - v_{n}^{'})},
\end{align}
\end{shrinkeq}
where $\mathbf{V}_{N - 1}^{n[-n]}$ represents the matrix $\mathbf{V}^{n}$ with its $n$-th row and column removed, and $\tilde{\mathbf{t}}_{N - 1}^{n[-n]}$ represents the vector $\tilde{\mathbf{t}}^{n}$ with its $n$-th entry $[\tilde{\mathbf{t}}^{n}]_{n}$ removed.

Based on the above procedure for $x_{n}$ ($n \in \mathcal{N}$), we are ready to compute the LMMSE-ISIC estimate $\hat{x}_{n}$ and the filtering bias $\mu_{n}$ under our proposed recursive scheme, which is stated by the following corollary.
\begin{cor}
Our proposed recursive scheme respectively computes the LMMSE-ISIC estimate $\hat{x}_{n}$ and the filtering bias $\mu_{n}$ by
\begin{shrinkeq}{-0.5ex}
\begin{align}
\hat{x}_{n} = \frac{v_{n}^{'}[\tilde{\mathbf{t}}^{'}]_{n} + \bar{x}_{n}^{'}(1 - \sigma^{2}\omega_{n}^{'})}{1 + \sigma^{2}\omega_{n}^{'}(v_{n}^{'} - 1)},
\end{align}
\end{shrinkeq}
and
\begin{shrinkeq}{-0.5ex}
\begin{align}
\mu_{n} = \frac{1 - \sigma^{2}\omega_{n}^{'}}{1 + \sigma^{2}\omega_{n}^{'}(v_{n}^{'} - 1)}.
\end{align}
\end{shrinkeq}
\end{cor}
\begin{proof}
As aforementioned, we can substitute $\bar{x}_{n}^{n} = 0$ and $v_{n}^{n} = 1$ shown in (25) and (26) into $\mathbf{Q}_{N}^{n}$ and $\tilde{\mathbf{t}}^{n}$ to obtain $\mathbf{Q}_{N}^{\hat{n}}$ and $\tilde{\mathbf{t}}^{\hat{n}}$, respectively. Hence, by recalling $\hat{x}_{n} = [\tilde{\mathbf{t}}^{\hat{n}}]_{n}$ given in \textbf{Proposition 1} and substituting (25) and (26) into (68), we obtain
\begin{shrinkeq}{-0.5ex}
\begin{align}
\hat{x}_{n} = [\tilde{\mathbf{t}}^{\hat{n}}]_{n} = \frac{v_{n}^{'}[\tilde{\mathbf{t}}^{'}]_{n} + \bar{x}_{n}^{'}(1 - \sigma^{2}\omega_{n}^{'})}{1 + \sigma^{2}\omega_{n}^{'}(v_{n}^{'} - 1)}.
\end{align}
\end{shrinkeq}
On the other hand, by recalling $\mu_{n} = 1 - \sigma^{2}[\mathbf{Q}_{N}^{\hat{n}}]_{n,n}$ given in \textbf{Proposition 2} and substituting (26) into (63), we obtain
\begin{shrinkeq}{-0.5ex}
\begin{align}
\mu_{n} = 1 - \sigma^{2}[\mathbf{Q}_{N}^{\hat{n}}]_{n,n} = \frac{1 - \sigma^{2}\omega_{n}^{'}}{1 + \sigma^{2}\omega_{n}^{'}(v_{n}^{'} - 1)},
\end{align}
\end{shrinkeq}
and thus \textbf{Corollary 1} has been proved.
\end{proof}

The above (63)-(70) are utilized to develop our proposed recursive low-complexity LMMSE-ISIC scheme, which is summarized in Algorithm 1.

\begin{algorithm*}[!t]
\caption{Proposed Recursive Low-Complexity LMMSE-ISIC Scheme}
\begin{algorithmic}[1]
\STATE Initialize $\bar{\mathbf{x}} = [\bar{x}_{1}, \cdots, \bar{x}_{N}]^{\mathsf{T}} \leftarrow \bm{0}_{N}$, $\mathbf{V} = \mathsf{Diag}([v_{1}, \cdots, v_{N}]^{\mathsf{T}}) \leftarrow \mathbf{I}_{N}$, $\mathbf{Q} \leftarrow (\mathbf{H}^{\mathsf{H}}\mathbf{H} + \sigma^{2}\mathbf{I}_{N})^{-1}$, and $\tilde{\mathbf{t}} \leftarrow \mathbf{Q}\mathbf{H}^{\mathsf{H}}\mathbf{y}$;
\WHILE{no iteration-stopping criteria are satisfied}
\FOR{$n = 1$ to $N$}
\STATE Compute $\hat{x}_{n} \leftarrow \frac{v_{n}[\tilde{\mathbf{t}}]_{n} + \bar{x}_{n}(1 - \sigma^{2}\omega_{n})}{1 + \sigma^{2}\omega_{n}(v_{n} - 1)}$ and $\mu_{n} \leftarrow \frac{1 - \sigma^{2}\omega_{n}}{1 + \sigma^{2}\omega_{n}(v_{n} - 1)}$, where $\omega_{n}$ represents $[\mathbf{Q}]_{n,n}$;
\STATE Compute $\eta_{n}^{2} \leftarrow \mu_{n}(1 - \mu_{n})$, and $P_{n}(x) \leftarrow \frac{\psi_{n}(x)}{\sum_{x' \in \mathcal{X}}\psi_{n}(x')}$ with $\psi_{n}(x) \leftarrow \mathsf{exp}(\frac{-|\hat{x}_{n} - \mu_{n}x|^{2}}{\eta_{n}^{2}})$ for all $x \in \mathcal{X}$;
\STATE Store the current $\omega_{n}$, $\bar{x}_{n}$ and $v_{n}$ for subsequent calculations, which are denoted as $\omega_{n}^{'}$, $\bar{x}_{n}^{'}$ and $v_{n}^{'}$, respectively;
\STATE Update $\bar{x}_{n} \leftarrow \sum_{x \in \mathcal{X}}xP_{n}(x)$ and $v_{n} \leftarrow \sum_{x \in \mathcal{X}}|x - \bar{x}_{n}|^{2}P_{n}(x)$;
\STATE Update $\tilde{\mathbf{t}}$ by $\tilde{\mathbf{t}}_{N - 1}^{[-n]} \leftarrow \tilde{\mathbf{t}}_{N - 1}^{[-n]} + \sigma^{2}\sqrt{v_{n}^{'}}(\mathbf{V}_{N - 1}^{[-n]})^{-\frac{1}{2}}[\mathbf{Q}]_{:,n}^{[-n]}\frac{\bar{x}_{n} - \bar{x}_{n}^{'} + (v_{n} - v_{n}^{'})[\tilde{\mathbf{t}}]_{n}}{v_{n} - \sigma^{2}\omega_{n}^{'}(v_{n} - v_{n}^{'})}$ and $[\tilde{\mathbf{t}}]_{n} \leftarrow \frac{v_{n}^{'}[\tilde{\mathbf{t}}]_{n} + (\bar{x}_{n} - \bar{x}_{n}^{'})(\sigma^{2}\omega_{n}^{'} - 1)}{v_{n} - \sigma^{2}\omega_{n}^{'}(v_{n} - v_{n}^{'})}$;
\STATE Compute $\omega_{n} \leftarrow \frac{\omega_{n}^{'}v_{n}^{'}}{v_{n} + \sigma^{2}\omega_{n}^{'}(v_{n}^{'} - v_{n})}$, and update $\mathbf{Q}$ by $\mathbf{Q}_{N - 1}^{[-n]} \leftarrow \mathbf{Q}_{N - 1}^{[-n]} + \frac{\omega_{n}v_{n} - \omega_{n}^{'}v_{n}^{'}}{(\omega_{n}^{'})^{2}v_{n}^{'}}[\mathbf{Q}]_{:,n}^{[-n]}([\mathbf{Q}]_{:,n}^{[-n]})^{\mathsf{H}}$, $[\mathbf{Q}]_{:,n}^{[-n]} \leftarrow $ $\frac{\omega_{n}}{\omega_{n}^{'}}\sqrt{\frac{v_{n}}{v_{n}^{'}}}[\mathbf{Q}]_{:,n}^{[-n]}$, $[\mathbf{Q}]_{n,:}^{[-n]} \leftarrow ([\mathbf{Q}]_{:,n}^{[-n]})^{\mathsf{H}}$, and $[\mathbf{Q}]_{n,n} \leftarrow \omega_{n}$;
\ENDFOR
\ENDWHILE
\STATE Compute $\check{x}_{n} \leftarrow \mathsf{arg}\max_{x \in \mathcal{X}}P_{n}(x)$ for $n \in \mathcal{N}$.
\end{algorithmic}
\end{algorithm*}

\vspace{-3mm}
\subsection{Complexity and Memory Requirement}

In this subsection, we compare the existing low-complexity LMMSE-ISIC scheme [24] and our proposed recursive low-complexity LMMSE-ISIC scheme by theoretical analysis, including complexity and memory requirement.

\subsubsection{Complexity}

Take the complexity analysis for the existing and our proposed low-complexity schemes into account first, and we will use $\langle j \rangle$ to denote the computational complexity of $j$ complex multiplications and $j$ complex additions. Besides, only the dominant complexities will be counted for simplicity.

For initialization, it can be observed that its complexity is dominated by computing the matrix product $\mathbf{H}^{\mathsf{H}}\mathbf{H}$ and the inverse $(\mathbf{H}^{\mathsf{H}}\mathbf{H} + \sigma^{2}\mathbf{I}_{N})^{-1}$ for both the existing and proposed low-complexity schemes, which require the complexities of $\langle \frac{1}{2}MN^{2} \rangle$ and $\langle \frac{1}{2}N^{3} \rangle$, respectively. Hence, the complexity of initialization for both the existing and proposed low-complexity schemes has an order of
\begin{shrinkeq}{-0.5ex}
\begin{align}
\langle \frac{1}{2}MN^{2} + \frac{1}{2}N^{3} \rangle.
\end{align}
\end{shrinkeq}

During the procedure for $x_{n}$ ($n \in \mathcal{N}$) in each iteration, the dominant complexity of the existing low-complexity scheme results from the calculation of $\mathbf{G}$, the calculation of $\mathbf{f}_{\mathrm{PR},n}$, and the calculation of $\tilde{\mathbf{y}}$ [24], which require the complexities of $\langle 2N^{2} \rangle$$^{\text{3}}$\footnote{$^{\text{3}}$Note $\mathbf{G} \triangleq (\mathbf{H}^{\mathsf{H}}\mathbf{H}\mathbf{V} + \sigma^{2}\mathbf{I}_{N})^{-1}$ shown in Sec. II-C is not a Hermitian matrix, since $\mathbf{H}^{\mathsf{H}}\mathbf{H}\mathbf{V}$ is not Hermitian because usually $(\mathbf{H}^{\mathsf{H}}\mathbf{H}\mathbf{V})^{\mathsf{H}} = \mathbf{V}\mathbf{H}^{\mathsf{H}}\mathbf{H} \neq \mathbf{H}^{\mathsf{H}}\mathbf{H}\mathbf{V}$.}, $\langle MN \rangle$, and $\langle MN \rangle$, respectively. Hence, in each iteration with $N$ procedures, the existing low-complexity scheme requires the dominant complexity of
\begin{shrinkeq}{-0.5ex}
\begin{align}
N(\langle 2N^{2} \rangle + \langle MN \rangle + \langle MN \rangle) = \langle 2N^{3} + 2MN^{2} \rangle.
\end{align}
\end{shrinkeq}
Instead, in each iteration with $N$ procedures, our proposed recursive low-complexity scheme only requires the dominant complexity of
\begin{shrinkeq}{-0.5ex}
\begin{align}
N\langle \frac{1}{2}N^{2} \rangle = \langle \frac{1}{2}N^{3} \rangle,
\end{align}
\end{shrinkeq}
which results from the update of the Hermitian matrix $\mathbf{Q}$ shown in (64) with the complexity of $\langle \frac{1}{2}N^{2} \rangle$$^{\text{4}}$\footnote{$^{\text{4}}$Only the triangular part of the Hermitian matrix $\mathbf{Q}$ needs to be computed.}.

According to (73)-(75), it can be observed that the speedup of our proposed low-complexity scheme over the existing counterpart is given by
\begin{shrinkeq}{-0.5ex}
\begin{align}
\frac{2N^{3} + 2MN^{2}}{\frac{1}{2}N^{3}} = \frac{4N + 4M}{N} \geq 8
\end{align}
\end{shrinkeq}
when comparing the complexity of each iteration, while the speedup becomes
\begin{shrinkeq}{-0.5ex}
\begin{align}
&\frac{\frac{1}{2}MN^{2} + \frac{1}{2}N^{3} + K(2N^{3} + 2MN^{2})}{\frac{1}{2}MN^{2} + \frac{1}{2}N^{3} + K(\frac{1}{2}N^{3})} \notag\\
&= \frac{1 + 4K}{1 + \frac{K}{1 + \frac{M}{N}}} \geq \frac{1 + 4K}{1 + \frac{K}{2}} = 1 + \frac{7}{1 + \frac{2}{K}}
\end{align}
\end{shrinkeq}
when comparing the total complexity including initialization and $K$ iterations, where $M \geq N$ is utilized. Furthermore, the minimum speedups given by (76) and (77) will be achieved with $M = N$. It is worth noting that although the existing low-complexity scheme successfully replaces the direct matrix inversion with the vector-based operations, it still needs the matrix-vector multiplications for computing $\mathbf{G}$ and $\mathbf{f}_{\mathrm{PR},n}$. In contrast, our proposed recursive scheme does not need any matrix-vector multiplications, and only needs the vector-vector multiplications to obtain the Hermitian matrix $\mathbf{Q}$ given in (64), which thus has a much lower complexity, especially when $M$ is sufficiently large.

\subsubsection{Memory Requirement}

Next, we analyze the memory requirements of the existing and our proposed low-complexity scheme, where we assume that one memory unit denotes the memory size to store one real number, and we only count the dominant memory units for simplicity. Accordingly, we only consider the memories for matrices.

It can be observed that during the procedure for $x_{n}$ in each iteration, the existing low-complexity scheme requires $N^{2}$, $2N^{2}$, and $2MN$ memory units to store the Hermitian matrix $\mathbf{H}^{\mathsf{H}}\mathbf{H}$, and the non-Hermitian matrices $\mathbf{G}$ and $\mathbf{H}$, respectively [24]. Instead, our proposed recursive low-complexity scheme only requires $N^{2}$ memory units to store the Hermitian matrix $\mathbf{Q}$. Then, compared to the existing low-complexity scheme that requires
\begin{shrinkeq}{-0.5ex}
\begin{align}
N^{2} + 2N^{2} + 2MN = 3N^{2} + 2MN
\end{align}
\end{shrinkeq}
memory units in total, our proposed scheme only requires
\begin{shrinkeq}{-0.5ex}
\begin{align}
\frac{N^{2}}{3N^{2} + 2MN} = \frac{N}{3N + 2M} \leq \frac{1}{5}
\end{align}
\end{shrinkeq}
memory units, and the maximum value of $\frac{1}{5}$ will be achieved when $M = N$. Therefore, we can conclude that our proposed recursive scheme exhibits a much lower memory requirement, especially when $M$ becomes larger.

\vspace{-2mm}
\section{Simulation Results}

In this section, we present extensive numerical results to demonstrate the benefit of our proposed recursive low-complexity LMMSE-ISIC scheme, which is superior over the existing low-complexity LMMSE-ISIC scheme [24].

\begin{figure}[!t]
\centering
\begin{subfigure}[c]{1\linewidth}
\centering
\includegraphics[scale=0.40]{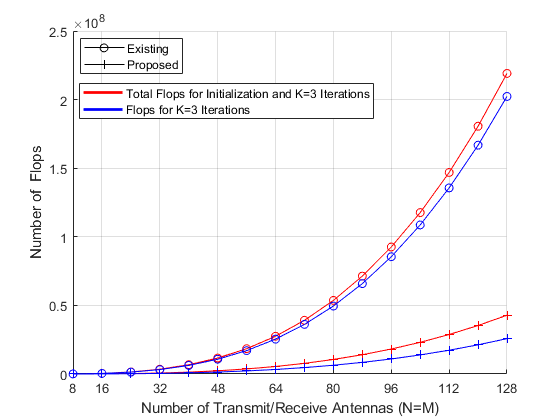}
\caption*{(a) Number of FLOPs.}
\end{subfigure}
\begin{subfigure}[c]{1\linewidth}
\centering
\includegraphics[scale=0.40]{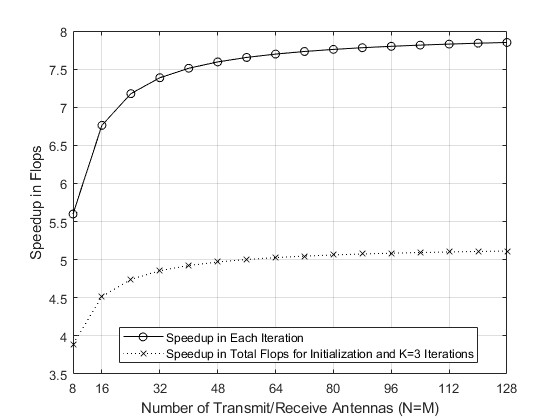}
\caption*{(b) Speedup in FLOPs.}
\end{subfigure}
\caption*{Fig. 1. Comparison of required FLOPs between the existing low-complexity LMMSE-ISIC scheme and our proposed recursive low-complexity LMMSE-ISIC scheme.}
\end{figure}

Fig. 1 compares the average floating-point operations (FLOPs) per data sample between the existing and our proposed low-complexity LMMSE-ISIC schemes$^{\text{5}}$\footnote{$^{\text{5}}$When counting the FLOPs of the statements in our MATLAB code, we utilize the fact that one complex multiplication and addition cost six and two FLOPs, respectively.}, where the number of transmit antenna is equal to that of receive antenna (i.e., $M = N$), and the number of ISIC iterations $K$ is selected as $3$ since it yields sufficiently fine BER performance (see Fig. 3 and corresponding statements). Fig. 1(a) shows the number of FLOPs per data sample for conducting the existing and our proposed low-complexity LMMSE-ISIC schemes, while Fig. 1(b) shows the corresponding speedups of our proposed scheme over the existing counterpart. As suggested by Fig. 1(a), with the growth of $M$, the required number of FLOPs for conducting the existing low-complexity LMMSE-ISIC scheme significantly increases, while increasing much slower under our proposed scheme. The reason lies in that, as aforementioned, the existing low-complexity LMMSE-ISIC scheme invokes matrix-vector multiplications in each iteration. Instead, our proposed scheme only invokes vector-vector multiplications in each iteration, which has much lower complexity, especially when $M$ becomes larger. Besides, it can be seen in Fig. 1(b) that our proposed low-complexity LMMSE-ISIC scheme reduces the complexity by at least $82.14\%$ in each iteration, while by at least $74.36\%$ in total for initialization and $K = 3$ iterations compared to the existing counterpart. Furthermore, with $M$ increasing, our proposed scheme yields much more speedups over the existing scheme, which approach the theoretical upper-bounds given in (76) and (77) when $M$ is modestly large. These phenomena verify the unique low-complexity feature of our proposed recursive LMMSE-ISIC scheme, which yields much lower complexity than the existing low-complexity scheme.

\begin{figure}[!t]
\centering
\includegraphics[scale=0.40]{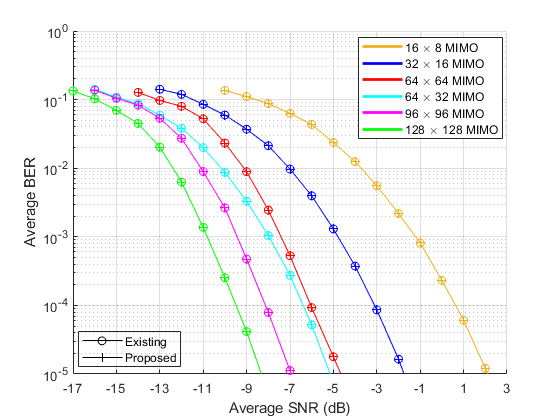}
\caption*{Fig. 2. Average uncoded BER performance v.s. antenna configuration, with 4-QAM modulation and $K = 3$.}
\end{figure}

Fig. 2 illustrates the average uncoded BER performance under different antenna configurations, where both the existing and our proposed low-complexity LMMSE-ISIC schemes are considered, and we choose 4-QAM modulation and $K = 3$. As reflected by this figure, the BER performance is better with $M$ and $N$ proportionally increasing. For instance, to achieve the same BER of $10^{-4}$, the required SNR is less than $-3$ dB for $32 \times 16$ MIMO, while larger than $0$ dB for $16 \times 8$ MIMO. Besides, it can be observed from this figure that $64 \times 32$ MIMO exhibits better BER performance than $64 \times 64$ MIMO, since much more data need to be uncoded for the latter case. Most importantly, our proposed low-complexity LMMSE-ISIC scheme yields identical BER performance with that obtained by the existing scheme, but with a significantly much lower complexity demonstrated by Fig. 1.

\begin{figure}[!t]
\centering
\includegraphics[scale=0.40]{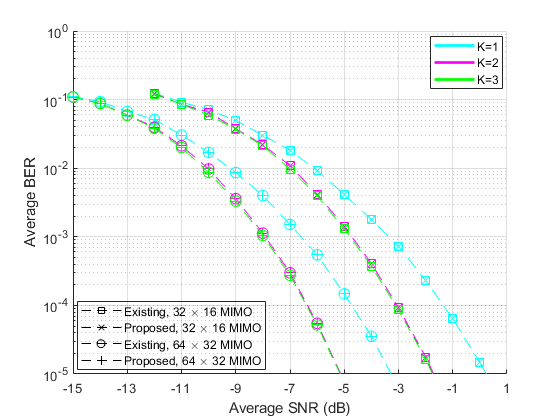}
\caption*{Fig. 3. Average uncoded BER performance v.s. number of ISIC iterations, with 4-QAM modulation.}
\end{figure}

\begin{figure}[!t]
\centering
\includegraphics[scale=0.40]{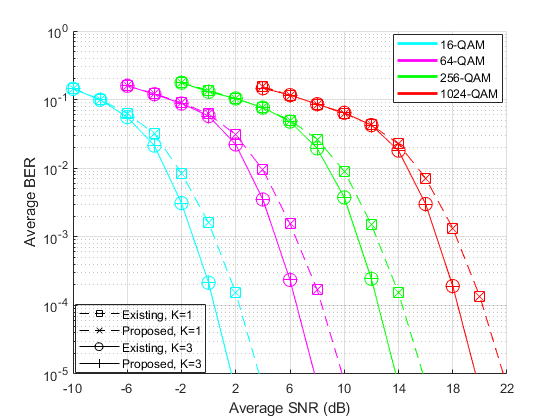}
\caption*{Fig. 4. Average uncoded BER performance v.s. modulation order, with $64 \times 32$ MIMO.}
\end{figure}

Fig. 3 plots the average uncoded BER performance under different number of ISIC iterations, where we again consider both the existing and our proposed low-complexity LMMSE-ISIC schemes, and 4-QAM modulation is adopted. As shown by this figure, the BER performance improves when $M$ and $N$ proportionally increase. Besides, the BER performance achieved by our proposed low-complexity LMMSE-ISIC scheme coincides that yielded by the existing low-complexity LMMSE-ISIC scheme, which, however, exhibits significantly much more lower complexity than the latter. In addition, intuitively the larger the number of ISIC iterations $K$ is, the better the BER performance is, and we can observe from Fig. 3 that $K = 1$ exhibits the worst BER performance compared to $K = 2$ and $K = 3$. Nevertheless, compared to $K = 2$, the improvement of BER performance resulting from $K = 3$ is almost negligible. Hence, $K = 3$ can be regarded as the case whose BER performance is adequately close to the best among different $K$'s, and we can select $3$ as the maximal number of iterations to avoid unnecessary computations, which can effectively strike the balance between performance and complexity.

Fig. 4 depicts the average uncoded BER performance under different modulation orders, where we consider a $64 \times 32$ MIMO under different low-complexity LMMSE-ISIC schemes and different number of ISIC iterations. It can be observed from this figure that, with a higher modulation order, the BER performance degrades, due to the fact that given the same SNR, the points on the constellation diagram will be more densely packed when the modulation order increases, i.e., the Euclidean distance between any two points shrinks, which undermines the anti-interference capability of considered system. Besides, the BER performances achieved by the existing and our proposed low-complexity LMMSE-ISIC schemes are identical, while the complexity of the latter is significantly much lower than that of the former (recall Fig. 1). In addition, the BER performance can be enhanced by increasing the number of iterations $K$, as shown by the curves associated with $K = 1$ and $K = 3$ in Fig. 4, respectively. Notably, the BER performance is sufficiently satisfactory when the number of iterations $K$ is chosen as $3$, which remarkably outperforms the case of $K = 1$ with only slight complexity increasement (recall Fig. 3).

\vspace{-2mm}
\section{Conclusion}

In this paper, we have presented a novel reordered description of the LMMSE-ISIC detection process for MIMO based on the equivalent channel matrix. We have obtained a critical observation that the detection process for LMMSE-ISIC is the extension of that for HD-OSIC by comparing the both according to the above reordered description. Then, we have proposed a recursive scheme for LMMSE-ISIC via extending that for HD-OSIC. Our proposed recursive low-complexity LMMSE-ISIC scheme is based on the Hermitian inverse matrix and the uniform symbol estimate vector, which are updated very efficiently, and applied to compute the LMMSE-ISIC estimate and the MMSE filtering bias. Both theoretical analyses and simulation results have demonstrated the advantages of our proposed low-complexity scheme. Specifically, according to theoretical analyses, compared to the existing counterpart, our proposed scheme only requires at most $\frac{1}{8}$ computations and $\frac{1}{5}$ memory units in each iteration, while requiring at most $\frac{1 + \frac{K}{2}}{1 + 4K}$ computations when taking initialization and $K$ iterations into account. On the other hand, it has been verified by simulations that the existing and our proposed low-complexity LMMSE-ISIC schemes can exhibit identical BER performance, however, the complexity of the latter is significantly much lower than that of the former. Interestingly, only $K = 3$ ISIC iterations are needed to yield sufficiently fine BER performance, which is close to the best among different $K$'s, and hence the balance between performance and complexity can be achieved.

Besides the recursive scheme for HD-OSIC that has been extended to develop our proposed solution in this paper, the square-root scheme, which is another typical scheme for HD-OSIC, can also be extended to design that for LMMSE-ISIC. Particularly, the complexity of the square-root LMMSE-ISIC detector may be lower than our proposed recursive counterpart, hence, it is meaningful to investigate this scheme, which is left to our future works.

\vspace{-2mm}
\appendix

\subsection{Derivation of (55)-(57)}

Define the inverse matrix of $\mathbf{Q}_{n}^{l}$ given in (43) as
\begin{shrinkeq}{-0.5ex}
\begin{align}
\mathbf{R}_{n}^{l} \triangleq (\mathbf{Q}_{n}^{l})^{-1} = (\tilde{\mathbf{H}}_{n}^{l})^{\mathsf{H}}\tilde{\mathbf{H}}_{n}^{l} + \sigma^{2}\mathbf{I}_{n}.
\end{align}
\end{shrinkeq}
Then, by substituting $\tilde{\mathbf{H}}_{n}^{l} = [\tilde{\mathbf{H}}_{n - 1}^{l}, \tilde{\mathbf{h}}_{n}^{l}]$ into the above (80), we have
\begin{shrinkeq}{-0.5ex}
\begin{align}
\mathbf{R}_{n}^{l} = \begin{bmatrix}
\mathbf{R}_{n - 1}^{l} & \ddot{\mathbf{r}}_{n}^{l}\\
(\ddot{\mathbf{r}}_{n}^{l})^{\mathsf{H}} & \gamma_{n}^{l}\\
\end{bmatrix}
\end{align}
\end{shrinkeq}
with
\begin{shrinkeq}{-0.5ex}
\begin{align}
\mathbf{R}_{n - 1}^{l} &= (\tilde{\mathbf{H}}_{n - 1}^{l})^{\mathsf{H}}\tilde{\mathbf{H}}_{n - 1}^{l} + \sigma^{2}\mathbf{I}_{n - 1}, \\
\ddot{\mathbf{r}}_{n}^{l} &= (\tilde{\mathbf{H}}_{n - 1}^{l})^{\mathsf{H}}\tilde{\mathbf{h}}_{n}^{l}, \\
\gamma_{n}^{l} &= (\tilde{\mathbf{h}}_{n}^{l})^{\mathsf{H}}\tilde{\mathbf{h}}_{n}^{l} + \sigma^{2}.
\end{align}
\end{shrinkeq}
Next, by substituting $\tilde{\mathbf{H}}^{N} = [\tilde{\mathbf{H}}_{N - 1}^{'}, \sqrt{\frac{v_{N}^{N}}{v_{N}^{'}}}\tilde{\mathbf{h}}_{N}^{'}]$ shown in (39) into (80) with $l = n = N$, we obtain$^{\text{6}}$\footnote{$^{\text{6}}$Note that the following $\mathbf{R}_{N - 1}^{'}$, $\ddot{\mathbf{r}}_{N}^{'}$, and $\gamma_{N}^{'}$ are the shorthand expressions for $\mathbf{R}_{N - 1}^{\mathsf{o}(N - 1)}$, $\ddot{\mathbf{r}}_{N}^{\mathsf{o}(N - 1)}$, and $\gamma_{N}^{\mathsf{o}(N - 1)}$, respectively.}
\begin{shrinkeq}{-0.5ex}
\begin{align}
\mathbf{R}_{N}^{N} = \begin{bmatrix}
\mathbf{R}_{N - 1}^{'} & \sqrt{\frac{v_{N}^{N}}{v_{N}^{'}}}\tilde{\mathbf{H}}_{N - 1}^{'\mathsf{H}}\tilde{\mathbf{h}}_{N}^{'}\\
\sqrt{\frac{v_{N}^{N}}{v_{N}^{'}}}\tilde{\mathbf{h}}_{N}^{'\mathsf{H}}\tilde{\mathbf{H}}_{N - 1}^{'} & \frac{v_{N}^{N}}{v_{N}^{'}}\tilde{\mathbf{h}}_{N}^{'\mathsf{H}}\tilde{\mathbf{h}}_{N}^{'} + \sigma^{2}\\
\end{bmatrix},
\end{align}
\end{shrinkeq}
where $\mathbf{R}_{N - 1}^{'} = \tilde{\mathbf{H}}_{N - 1}^{'\mathsf{H}}\tilde{\mathbf{H}}_{N - 1}^{'} + \sigma^{2}\mathbf{I}_{N - 1}$ due to (82) with $l = \mathsf{o}(N - 1)$ and $n = N$. After substituting (83) and (84) (equivalent to $(\tilde{\mathbf{h}}_{n}^{l})^{\mathsf{H}}\tilde{\mathbf{h}}_{n}^{l} = \gamma_{n}^{l} - \sigma^{2}$) with $l = \mathsf{o}(N - 1)$ and $n = N$ into (85), followed by comparing (85) and (81) with $l = n = N$, the sub-blocks within (81) can be given by
\begin{shrinkeq}{-0.5ex}
\begin{align}
\mathbf{R}_{N - 1}^{N} &= \mathbf{R}_{N - 1}^{'}, \\
\ddot{\mathbf{r}}_{N}^{N} &= \sqrt{\frac{v_{N}^{N}}{v_{N}^{'}}}\ddot{\mathbf{r}}_{N}^{'}, \\
\gamma_{N}^{N} &= (\gamma_{N}^{'} - \sigma^{2})\frac{v_{N}^{N}}{v_{N}^{'}} + \sigma^{2}.
\end{align}
\end{shrinkeq}

According to the lemma for inverting a partitioned matrix [32], $\ddot{\mathbf{Q}}_{N - 1}^{N}$, $\ddot{\mathbf{q}}_{N}^{N}$, and $\omega_{N}^{N}$ within $\mathbf{Q}_{N}^{N}$ given in (54) can be calculated as
\begin{shrinkeq}{-0.5ex}
\begin{align}
\omega_{N}^{N} &= (\gamma_{N}^{N} - (\ddot{\mathbf{r}}_{N}^{N})^{\mathsf{H}}\mathbf{Q}_{N - 1}^{N}\ddot{\mathbf{r}}_{N}^{N})^{-1}, \\
\ddot{\mathbf{q}}_{N}^{N} &= -\omega_{N}^{N}\mathbf{Q}_{N - 1}^{N}\ddot{\mathbf{r}}_{N}^{N}, \\
\ddot{\mathbf{Q}}_{N - 1}^{N} &= \mathbf{Q}_{N - 1}^{N} + (\omega_{N}^{N})^{-1}\ddot{\mathbf{q}}_{N}^{N}(\ddot{\mathbf{q}}_{N}^{N})^{\mathsf{H}},
\end{align}
\end{shrinkeq}
where $\mathbf{Q}_{N - 1}^{N}$ can be further casted as
\begin{shrinkeq}{-0.5ex}
\begin{align}
\mathbf{Q}_{N - 1}^{N} = \mathbf{Q}_{N - 1}^{'},
\end{align}
\end{shrinkeq}
by substituting $\mathbf{R}_{n}^{l} \triangleq (\mathbf{Q}_{n}^{l})^{-1}$ given in (80) into (86). It can be observed from (91) (with each term's superscript $(\cdot)^{N}$ being replaced by the superscript $(\cdot)^{'}$) that $\mathbf{Q}_{N - 1}^{'} = \ddot{\mathbf{Q}}_{N - 1}^{'} - \omega_{N}^{'-1}\ddot{\mathbf{q}}_{N}^{'}(\ddot{\mathbf{q}}_{N}^{'})^{\mathsf{H}}$, and based on (92), we have
\begin{shrinkeq}{-0.5ex}
\begin{align}
\mathbf{Q}_{N - 1}^{N} = \ddot{\mathbf{Q}}_{N - 1}^{'} - \omega_{N}^{'-1}\ddot{\mathbf{q}}_{N}^{'}\ddot{\mathbf{q}}_{N}^{'\mathsf{H}}.
\end{align}
\end{shrinkeq}

To derive (55) and (56), we can substitute (87), (88) and (92) into (89) and (90) to obtain
\begin{shrinkeq}{-0.5ex}
\begin{align}
\omega_{N}^{N} = \bigg((\gamma_{N}^{'} - \ddot{\mathbf{r}}_{N}^{'\mathsf{H}}\mathbf{Q}_{N - 1}^{'}\ddot{\mathbf{r}}_{N}^{'} - \sigma^{2})\frac{v_{N}^{N}}{v_{N}^{'}} + \sigma^{2} \bigg)^{-1},
\end{align}
\end{shrinkeq}
and
\begin{shrinkeq}{-0.5ex}
\begin{align}
\ddot{\mathbf{q}}_{N}^{N} = -\frac{\omega_{N}^{N}}{\omega_{N}^{'}}\sqrt{\frac{v_{N}^{N}}{v_{N}^{'}}}\omega_{N}^{'}\mathbf{Q}_{N - 1}^{'}\ddot{\mathbf{r}}_{N}^{'},
\end{align}
\end{shrinkeq}
respectively, followed by substituting (89) and (90) (with each term's superscript $(\cdot)^{N}$ being replaced by the superscript $(\cdot)^{'}$) into (94) and (95), respectively. To obtain (57), we can substitute (56) and (93) into (91), and thus \textbf{Proposition 3} has been proved.

\subsection{Derivation of (58) and (59)}

We first recast the uniform symbol estimate vector $\tilde{\mathbf{t}}^{l} = (\mathbf{V}^{l})^{-\frac{1}{2}}\mathbf{Q}_{N}^{l}(\tilde{\mathbf{H}}^{l})^{\mathsf{H}}\tilde{\mathbf{y}}^{l}$ given in (45) as
\begin{shrinkeq}{-0.5ex}
\begin{align}
\tilde{\mathbf{t}}^{l} = (\mathbf{V}^{l})^{-\frac{1}{2}}\mathbf{t}^{l}, \; l \in \{ \mathsf{o}(n - 1), \hat{n}, n \},
\end{align}
\end{shrinkeq}
where $\mathbf{t}^{l} \triangleq \mathbf{Q}_{N}^{l}(\tilde{\mathbf{H}}^{l})^{\mathsf{H}}\tilde{\mathbf{y}}^{l}$ can be further written as
\begin{shrinkeq}{-0.5ex}
\begin{align}
\mathbf{t}^{l} = \mathbf{Q}_{N}^{l}\mathbf{z}^{l}
\end{align}
\end{shrinkeq}
with
\begin{shrinkeq}{-0.5ex}
\begin{align}
\mathbf{z}^{l} \triangleq (\tilde{\mathbf{H}}^{l})^{\mathsf{H}}\tilde{\mathbf{y}}^{l}.
\end{align}
\end{shrinkeq}
Then, to achieve the proof of \textbf{Proposition 4}, we derive the efficient procedures for updating $\mathbf{z}^{l}$ and $\mathbf{t}^{l}$ successively, and finally obtain (58) and (59). Note in the following, we will utilize $\mathbf{z}^{'}$ to represent $\mathbf{z}^{\mathsf{o}(N - 1)}$ for brevity.

Let us take the efficient update for $\mathbf{z}^{l}$ into account. According to $\mathbf{h}_{n} = \frac{\tilde{\mathbf{h}}_{n}^{l}}{\sqrt{v_{n}^{l}}}$ given in (41), the equalities $\mathbf{h}_{N} = \frac{\tilde{\mathbf{h}}_{N}^{'}}{\sqrt{v_{N}^{'}}}$ and $\mathbf{h}_{N} = \frac{\tilde{\mathbf{h}}_{N}^{N}}{\sqrt{v_{N}^{N}}}$ hold, which implies
\begin{shrinkeq}{-0.5ex}
\begin{align}
\frac{\tilde{\mathbf{h}}_{N}^{'}}{\sqrt{v_{N}^{'}}} = \frac{\tilde{\mathbf{h}}_{N}^{N}}{\sqrt{v_{N}^{N}}}.
\end{align}
\end{shrinkeq}
By substituting (99) into $\tilde{\mathbf{y}}^{N} = \tilde{\mathbf{y}}^{'} - \frac{\bar{x}_{N}^{N} - \bar{x}_{N}^{'}}{\sqrt{v_{N}^{'}}}\tilde{\mathbf{h}}_{N}^{'}$ given in (38), we have
\begin{shrinkeq}{-0.5ex}
\begin{align}
\tilde{\mathbf{y}}^{N} = \tilde{\mathbf{y}}^{'} - \frac{\bar{x}_{N}^{N} - \bar{x}_{N}^{'}}{\sqrt{v_{N}^{N}}}\tilde{\mathbf{h}}_{N}^{N}.
\end{align}
\end{shrinkeq}
Then, we substitute (100) into (98) with $l = N$ to obtain
\begin{shrinkeq}{-0.5ex}
\begin{align}
\mathbf{z}^{N} = (\tilde{\mathbf{H}}^{N})^{\mathsf{H}}\tilde{\mathbf{y}}^{'} - \frac{\bar{x}_{N}^{N} - \bar{x}_{N}^{'}}{\sqrt{v_{N}^{N}}}(\tilde{\mathbf{H}}^{N})^{\mathsf{H}}\tilde{\mathbf{h}}_{N}^{N}.
\end{align}
\end{shrinkeq}
Furthermore, by substituting $\tilde{\mathbf{H}}^{N} = [\tilde{\mathbf{H}}_{N - 1}^{'}, \sqrt{\frac{v_{N}^{N}}{v_{N}^{'}}}\tilde{\mathbf{h}}_{N}^{'}]$ given in (39) into $(\tilde{\mathbf{H}}^{N})^{\mathsf{H}}\tilde{\mathbf{y}}^{'}$ within (101), we have
\begin{shrinkeq}{-0.5ex}
\begin{align}
(\tilde{\mathbf{H}}^{N})^{\mathsf{H}}\tilde{\mathbf{y}}^{'} &= \begin{bmatrix}
\tilde{\mathbf{H}}_{N - 1}^{'\mathsf{H}}\\
\sqrt{\frac{v_{N}^{N}}{v_{N}^{'}}}\tilde{\mathbf{h}}_{N}^{'\mathsf{H}}\\
\end{bmatrix}\tilde{\mathbf{y}}^{'} = \begin{bmatrix}
\tilde{\mathbf{H}}_{N - 1}^{'\mathsf{H}}\tilde{\mathbf{y}}^{'}\\
\sqrt{\frac{v_{N}^{N}}{v_{N}^{'}}}\tilde{\mathbf{h}}_{N}^{'\mathsf{H}}\tilde{\mathbf{y}}^{'}\\
\end{bmatrix} \notag\\
&= \begin{bmatrix}
\mathbf{z}_{N - 1}^{'}\\
\sqrt{\frac{v_{N}^{N}}{v_{N}^{'}}}[\mathbf{z}^{'}]_{N}\\
\end{bmatrix},
\end{align}
\end{shrinkeq}
where (98) with $l = \mathsf{o}(N - 1)$ is utilized, and $\mathbf{z}_{N - 1}^{'}$ denotes the first $(N - 1)$ entries of $\mathbf{z}^{'}$. Finally, by substituting (102) into (101), we obtain
\begin{shrinkeq}{-0.5ex}
\begin{align}
\mathbf{z}^{N} = \begin{bmatrix}
\mathbf{z}_{N - 1}^{'}\\
\sqrt{\frac{v_{N}^{N}}{v_{N}^{'}}}[\mathbf{z}^{'}]_{N}\\
\end{bmatrix} - \frac{\bar{x}_{N}^{N} - \bar{x}_{N}^{'}}{\sqrt{v_{N}^{N}}}(\tilde{\mathbf{H}}^{N})^{\mathsf{H}}\tilde{\mathbf{h}}_{N}^{N}.
\end{align}
\end{shrinkeq}

Note that (103) requires to compute $(\tilde{\mathbf{H}}^{N})^{\mathsf{H}}\tilde{\mathbf{h}}_{N}^{N}$, which is relevant to the $N$-th column of $\mathbf{H}^{\mathsf{H}}\mathbf{H}$. To avoid storing $\mathbf{H}^{\mathsf{H}}\mathbf{H}$, we replace $\mathbf{z}^{l}$ with $\mathbf{t}^{l} = \mathbf{Q}_{N}^{l}\mathbf{z}^{l}$, i.e., (97), and update $\mathbf{t}^{l}$ by utilizing only the entries in $\mathbf{Q}_{N}^{'}$ and $\mathbf{Q}_{N}^{N}$. Specifically, by substituting (103) into (97) with $l = N$, we have
\begin{shrinkeq}{-0.5ex}
\begin{align}
\mathbf{t}^{N} = \xi - \frac{\bar{x}_{N}^{N} - \bar{x}_{N}^{'}}{\sqrt{v_{N}^{N}}}\mathbf{Q}_{N}^{N}(\tilde{\mathbf{H}}^{N})^{\mathsf{H}}\tilde{\mathbf{h}}_{N}^{N},
\end{align}
\end{shrinkeq}
where
\begin{shrinkeq}{-0.5ex}
\begin{align}
\xi \triangleq \mathbf{Q}_{N}^{N} \bigg[\mathbf{z}_{N - 1}^{'\mathsf{T}}, \sqrt{\frac{v_{N}^{N}}{v_{N}^{'}}}[\mathbf{z}^{'}]_{N} \bigg]^{\mathsf{T}}.
\end{align}
\end{shrinkeq}
Based on $\mathbf{Q}_{N}^{l}(\tilde{\mathbf{H}}^{l})^{\mathsf{H}}\tilde{\mathbf{H}}^{l} = \mathbf{I}_{N} - \sigma^{2}\mathbf{Q}_{N}^{l}$ given in (52), the $N$-th column of $\mathbf{Q}_{N}^{N}(\tilde{\mathbf{H}}^{N})^{\mathsf{H}}\tilde{\mathbf{H}}^{N}$ is given by
\begin{shrinkeq}{-0.5ex}
\begin{align}
\mathbf{Q}_{N}^{N}(\tilde{\mathbf{H}}^{N})^{\mathsf{H}}\tilde{\mathbf{h}}_{N}^{N} \!\!=\!\! \mathbf{e}_{N} \!-\! \sigma^{2}\mathbf{q}_{N}^{N} \!\!=\!\! [-\sigma^{2}(\ddot{\mathbf{q}}_{N}^{N})^{\mathsf{T}}, 1 \!-\! \sigma^{2}\omega_{N}^{N}]^{\mathsf{T}},
\end{align}
\end{shrinkeq}
where $\mathbf{e}_{n}$ ($n \in \mathcal{N}$) represents the $n$-th column of $\mathbf{I}_{N}$. Then, by substituting (106) into (104), $\mathbf{t}^{N}$ can be expressed as
\begin{shrinkeq}{-0.5ex}
\begin{align}
\mathbf{t}^{N} = \xi + \frac{\bar{x}_{N}^{N} - \bar{x}_{N}^{'}}{\sqrt{v_{N}^{N}}}[\sigma^{2}(\ddot{\mathbf{q}}_{N}^{N})^{\mathsf{T}}, \sigma^{2}\omega_{N}^{N} - 1]^{\mathsf{T}}.
\end{align}
\end{shrinkeq}
On the other hand, by substituting $\mathbf{Q}_{N}^{'}$ that has the same structure as (54) into (97) with $l = \mathsf{o}(N - 1)$, we obtain
\begin{shrinkeq}{-0.5ex}
\begin{align}
\mathbf{t}^{'} &= \begin{bmatrix}
\mathbf{t}_{N - 1}^{'}\\
[\mathbf{t}^{'}]_{N}\\
\end{bmatrix} = \begin{bmatrix}
\ddot{\mathbf{Q}}_{N - 1}^{'} & \ddot{\mathbf{q}}_{N}^{'}\\
\ddot{\mathbf{q}}_{N}^{'\mathsf{H}} & \omega_{N}^{'}\\
\end{bmatrix}\mathbf{z}^{'} \notag\\
&= \begin{bmatrix}
\ddot{\mathbf{Q}}_{N - 1}^{'}\mathbf{z}_{N - 1}^{'} + \ddot{\mathbf{q}}_{N}^{'}[\mathbf{z}^{'}]_{N}\\
\ddot{\mathbf{q}}_{N}^{'\mathsf{H}}\mathbf{z}_{N - 1}^{'} + \omega_{N}^{'}[\mathbf{z}^{'}]_{N}\\
\end{bmatrix},
\end{align}
\end{shrinkeq}
and hence
\begin{shrinkeq}{-0.5ex}
\begin{align}
\ddot{\mathbf{Q}}_{N - 1}^{'}\mathbf{z}_{N - 1}^{'} &= \mathbf{t}_{N - 1}^{'} - \ddot{\mathbf{q}}_{N}^{'}[\mathbf{z}^{'}]_{N}, \\
\ddot{\mathbf{q}}_{N}^{'\mathsf{H}}\mathbf{z}_{N - 1}^{'} &= [\mathbf{t}^{'}]_{N} - \omega_{N}^{'}[\mathbf{z}^{'}]_{N}.
\end{align}
\end{shrinkeq}
By leveraging \textbf{Proposition 3} and after some manipulations, (105) can be rewritten as
\begin{shrinkeq}{-0.5ex}
\begin{align}
\xi &= \begin{bmatrix}
\ddot{\mathbf{Q}}_{N - 1}^{'} + \frac{\omega_{N}^{N}v_{N}^{N} - \omega_{N}^{'}v_{N}^{'}}{\omega_{N}^{'2}v_{N}^{'}}\ddot{\mathbf{q}}_{N}^{'}\ddot{\mathbf{q}}_{N}^{'\mathsf{H}} & \frac{\omega_{N}^{N}}{\omega_{N}^{'}}\sqrt{\frac{v_{N}^{N}}{v_{N}^{'}}}\ddot{\mathbf{q}}_{N}^{'}\\
\frac{\omega_{N}^{N}}{\omega_{N}^{'}}\sqrt{\frac{v_{N}^{N}}{v_{N}^{'}}}\ddot{\mathbf{q}}_{N}^{'\mathsf{H}} & \omega_{N}^{N}\\
\end{bmatrix} \notag\\
& \times \bigg[\mathbf{z}_{N - 1}^{'\mathsf{T}}, \sqrt{\frac{v_{N}^{N}}{v_{N}^{'}}}[\mathbf{z}^{'}]_{N} \bigg]^{\mathsf{T}} \notag\\
&= \begin{bmatrix}
\mathbf{t}_{N - 1}^{'} + [\mathbf{t}^{'}]_{N}\frac{\omega_{N}^{N}v_{N}^{N} - \omega_{N}^{'}v_{N}^{'}}{\omega_{N}^{'2}v_{N}^{'}}\ddot{\mathbf{q}}_{N}^{'}\\
\frac{\omega_{N}^{N}}{\omega_{N}^{'}}\sqrt{\frac{v_{N}^{N}}{v_{N}^{'}}}[\mathbf{t}^{'}]_{N}\\
\end{bmatrix},
\end{align}
\end{shrinkeq}
where (109) and (110) are utilized. Then, by substituting (111) into (107), $\mathbf{t}^{N}$ can be recasted as
\begin{shrinkeq}{-0.5ex}
\begin{align}
\mathbf{t}^{N} = \begin{bmatrix}
\mathbf{t}_{N - 1}^{N}\\
[\mathbf{t}^{N}]_{N}\\
\end{bmatrix} &= \begin{bmatrix}
\mathbf{t}_{N - 1}^{'} + [\mathbf{t}^{'}]_{N}\frac{\omega_{N}^{N}v_{N}^{N} - \omega_{N}^{'}v_{N}^{'}}{\omega_{N}^{'2}v_{N}^{'}}\ddot{\mathbf{q}}_{N}^{'}\\
\frac{\omega_{N}^{N}}{\omega_{N}^{'}}\sqrt{\frac{v_{N}^{N}}{v_{N}^{'}}}[\mathbf{t}^{'}]_{N}\\
\end{bmatrix} \notag\\
& + \frac{\bar{x}_{N}^{N} - \bar{x}_{N}^{'}}{\sqrt{v_{N}^{N}}}\begin{bmatrix}
\sigma^{2}\ddot{\mathbf{q}}_{N}^{N}\\
\sigma^{2}\omega_{N}^{N} - 1\\
\end{bmatrix}.
\end{align}
\end{shrinkeq}
According to (112), we have
\begin{shrinkeq}{-0.5ex}
\begin{align}
\mathbf{t}_{N - 1}^{N} &= \mathbf{t}_{N - 1}^{'} + [\mathbf{t}^{'}]_{N}\frac{\omega_{N}^{N}v_{N}^{N} - \omega_{N}^{'}v_{N}^{'}}{\omega_{N}^{'2}v_{N}^{'}}\ddot{\mathbf{q}}_{N}^{'} \notag\\
& + \frac{\bar{x}_{N}^{N} - \bar{x}_{N}^{'}}{\sqrt{v_{N}^{N}}}\sigma^{2}\ddot{\mathbf{q}}_{N}^{N}.
\end{align}
\end{shrinkeq}
Then, by substituting $\ddot{\mathbf{q}}_{N} = \frac{\omega_{N}^{N}}{\omega_{N}^{'}}\sqrt{\frac{v_{N}^{N}}{v_{N}^{'}}}\ddot{\mathbf{q}}_{N}^{'}$ shown in (56) into (113), we obtain
\begin{shrinkeq}{-0.5ex}
\begin{align}
\mathbf{t}_{N - 1}^{N} &= \mathbf{t}_{N - 1}^{'} + [\mathbf{t}^{'}]_{N}\frac{\omega_{N}^{N}v_{N}^{N} - \omega_{N}^{'}v_{N}^{'}}{\omega_{N}^{'2}v_{N}^{'}}\ddot{\mathbf{q}}_{N}^{'} \notag\\
& + \sigma^{2}\omega_{N}^{N}\frac{\bar{x}_{N}^{N} - \bar{x}_{N}^{'}}{\omega_{N}^{'}\sqrt{v_{N}^{'}}}\ddot{\mathbf{q}}_{N}^{'}.
\end{align}
\end{shrinkeq}
For another, $[\mathbf{t}^{N}]_{N}$ can be given by
\begin{shrinkeq}{-0.5ex}
\begin{align}
[\mathbf{t}^{N}]_{N} = \frac{\omega_{N}^{N}}{\omega_{N}^{'}}\sqrt{\frac{v_{N}^{N}}{v_{N}^{'}}}[\mathbf{t}^{'}]_{N} + \frac{(\sigma^{2}\omega_{N}^{N} - 1)(\bar{x}_{N}^{N} - \bar{x}_{N}^{'})}{\sqrt{v_{N}^{N}}}.
\end{align}
\end{shrinkeq}
It can be observed from (114) and (115) that $\mathbf{t}^{l}$ is updated efficiently since the high-complexity matrix-vector multiplications in (104) and (105) are not needed any more.

Lastly, we proceed to derive (58) and (59) by leveraging the efficient procedure for updating $\mathbf{t}^{l}$ shown in (114) and (115). Before exposition, we first present some facts that are necessary for derivations. Specifically, based on $\tilde{\mathbf{t}}^{l} = (\mathbf{V}^{l})^{-\frac{1}{2}}\mathbf{t}^{l}$ given in (96), we have
\begin{shrinkeq}{-0.5ex}
\begin{align}
[\tilde{\mathbf{t}}^{N}]_{N} &= \frac{[\mathbf{t}^{N}]_{N}}{\sqrt{v_{N}^{N}}}, \\
[\tilde{\mathbf{t}}^{'}]_{N} &= \frac{[\mathbf{t}^{'}]_{N}}{\sqrt{v_{N}^{'}}}, \\
\tilde{\mathbf{t}}_{N - 1}^{N} &= (\mathbf{V}_{N - 1}^{N})^{-\frac{1}{2}}\mathbf{t}_{N - 1}^{N}, \\
\tilde{\mathbf{t}}_{N - 1}^{'} &= \mathbf{V}_{N - 1}^{'-\frac{1}{2}}\mathbf{t}_{N - 1}^{'}.
\end{align}
\end{shrinkeq}
Furthermore, it can be observed from (37) that $\mathbf{V}_{N - 1}^{N} = \mathbf{V}_{N - 1}^{'}$. Therefore, $\tilde{\mathbf{t}}_{N - 1}^{'}$ in (119) can be rewritten as
\begin{shrinkeq}{-0.5ex}
\begin{align}
\tilde{\mathbf{t}}_{N - 1}^{'} = (\mathbf{V}_{N - 1}^{N})^{-\frac{1}{2}}\mathbf{t}_{N - 1}^{'}.
\end{align}
\end{shrinkeq}
According to the above elaborations, we are ready to derive (58) and (59) in the sequel. Specifically, via multiplying (114) by $(\mathbf{V}_{N - 1}^{N})^{-\frac{1}{2}}$ and after some manipulations, we obtain
\begin{shrinkeq}{-0.5ex}
\begin{align}
&(\mathbf{V}_{N - 1}^{N})^{-\frac{1}{2}}\mathbf{t}_{N - 1}^{N} = (\mathbf{V}_{N - 1}^{N})^{-\frac{1}{2}}\mathbf{t}_{N - 1}^{'} + \frac{(\mathbf{V}_{N - 1}^{N})^{-\frac{1}{2}}\ddot{\mathbf{q}}_{N}^{'}}{\omega_{N}^{'}\sqrt{v_{N}^{'}}} \notag\\
& \times \bigg(\sqrt{v_{N}^{'}}[\mathbf{t}^{'}]_{N}\bigg(\frac{\omega_{N}^{N}v_{N}^{N}}{\omega_{N}^{'}v_{N}^{'}} - 1 \bigg) + \sigma^{2}\omega_{N}^{N}(\bar{x}_{N}^{N} - \bar{x}_{N}^{'}) \bigg).
\end{align}
\end{shrinkeq}
Then, by substituting (117), (118) and (120) into (121), we have
\begin{shrinkeq}{-0.5ex}
\begin{align}
\tilde{\mathbf{t}}_{N - 1}^{N} &= \tilde{\mathbf{t}}_{N - 1}^{'} + \frac{(\mathbf{V}_{N - 1}^{N})^{-\frac{1}{2}}\ddot{\mathbf{q}}_{N}^{'}}{\omega_{N}^{'}\sqrt{v_{N}^{'}}}\bigg([\tilde{\mathbf{t}}^{'}]_{N}\bigg(\frac{\omega_{N}^{N}}{\omega_{N}^{'}}v_{N}^{N} - v_{N}^{'} \bigg) \notag\\
& + \sigma^{2}\omega_{N}^{N}(\bar{x}_{N}^{N} - \bar{x}_{N}^{'}) \bigg).
\end{align}
\end{shrinkeq}
Finally, by substituting $\omega_{N}^{N} = \frac{\omega_{N}^{'}v_{N}^{'}}{v_{N}^{N} + \sigma^{2}\omega_{N}^{'}(v_{N}^{'} - v_{N}^{N})}$ given in (55) into (122) and after some manipulations, (58) can be achieved. On the other hand, via multiplying (115) by $\frac{1}{\sqrt{v_{N}^{N}}}$ and after some manipulations, we obtain
\begin{shrinkeq}{-0.5ex}
\begin{align}
\frac{[\mathbf{t}^{N}]_{N}}{\sqrt{v_{N}^{N}}} = [\mathbf{t}^{'}]_{N}\frac{\omega_{N}^{N}}{\omega_{N}^{'}\sqrt{v_{N}^{'}}} + \frac{(\sigma^{2}\omega_{N}^{N} - 1)(\bar{x}_{N}^{N} - \bar{x}_{N}^{'})}{v_{N}^{N}}.
\end{align}
\end{shrinkeq}
Then, by substituting (116) and (117) into (123), we have
\begin{shrinkeq}{-0.5ex}
\begin{align}
[\tilde{\mathbf{t}}^{N}]_{N} = [\tilde{\mathbf{t}}^{'}]_{N}\frac{\omega_{N}^{N}}{\omega_{N}^{'}} + \frac{(\sigma^{2}\omega_{N}^{N} - 1)(\bar{x}_{N}^{N} - \bar{x}_{N}^{'})}{v_{N}^{N}}.
\end{align}
\end{shrinkeq}
Finally, by again substituting (55) into (124) and after some manipulations, (59) can be acquired, and thus the proof of \textbf{Proposition 4} has been completed.

\end{document}